\def\be{\begin{equation}}
\def\ee{\end{equation}}
\def\bea{\begin{eqnarray}}
\def\eea{\end{eqnarray}}

\documentclass[11pt]{article}

\usepackage{amsmath}
\usepackage{amssymb}

\usepackage{graphicx}

\usepackage{cite}
\usepackage{color} 
\usepackage{ifthen}  
\usepackage{longtable}
\usepackage{listing}

\usepackage{hyperref}
\usepackage{ulem}
\definecolor{lightyellow}{rgb}{1,1,.20}
\definecolor{lightblue}{rgb}{0.0,0.5,1}
\definecolor{lightred}{rgb}{1,0.6,0.6}

\newcommand\rsout{\bgroup\markoverwith
{\raisebox{.9ex}{\textcolor{blue}{\rule[-0.5ex]{2pt}{1.0pt}}}}\ULon}

\newcommand{\rev}[1]{\textcolor{black}{#1}}

\usepackage[labelfont=bf,labelsep=period,justification=raggedright]{caption}

\makeatletter
\renewcommand{\@biblabel}[1]{\quad#1.}
\makeatother

\date{}

\begin{document}
\renewcommand{\baselinestretch}{1.25}

\begin{flushleft}
{\Large
\textbf{RAId\_aPS: MS/MS analysis with multiple scoring functions and \\ 
\vspace*{4pt} spectrum-specific statistics}
}
\vspace*{7pt}\\
Gelio Alves$^{1}$, Aleksey Y. Ogurtsov$^{1}$  and                
Yi-Kuo Yu$^{1,\ast}$ \vspace*{4pt}
\\
\bf{1} National Center for Biotechnology Information, 
\mbox{National} Library of Medicine,\\ 
National Institutes of Health, Bethesda, MD 20894
\\
$\ast$ E-mail: \href{yyu@ncbi.nlm.nih.gov}{yyu@ncbi.nlm.nih.gov}
\end{flushleft}

\renewcommand{\baselinestretch}{1.35}
\section*{Abstract}
{\sf
Statistically meaningful comparison/combination of peptide identification results from various search methods 
  is impeded by the lack of a universal statistical standard. 
 Providing an $E$-value calibration protocol, we demonstrated earlier the feasibility 
 of translating either the score or heuristic $E$-value reported by any method into the 
 textbook-defined $E$-value, which may serve as the universal statistical standard. 
 This protocol, although robust, may lose spectrum-specific statistics and might 
 require a new calibration when changes in experimental setup occur. 
 To mitigate these issues, we developed a new MS/MS search tool, RAId\_aPS, that is able to  
 provide {\it spectrum-specific} $E$-values for {\it additive} scoring functions. 
 Given a selection of scoring functions out of RAId score, K-score, Hyperscore and XCorr,  
 RAId\_aPS generates the corresponding score histograms
 of {\it all possible} peptides using dynamic programming.  
 Using these score histograms to assign $E$-values enables a calibration-free 
 protocol for accurate significance assignment \rev{for each scoring function}. 
  RAId\_aPS features four different modes: (i) compute the total number of possible peptides for a given molecular mass range, 
  (ii) generate the score histogram given a MS/MS spectrum and a scoring function, 
  (iii) reassign $E$-values for a list of candidate peptides given a MS/MS spectrum and 
   the scoring functions chosen, and (iv) perform database searches using selected scoring functions.   
   In modes (iii) and (iv), RAId\_aPS is also capable of combining results from different scoring functions 
 using spectrum-specific statistics.\\
The web link is \href{http://www.ncbi.nlm.nih.gov/CBBresearch/Yu/raid_aps/index.html}{http://www.ncbi.nlm.nih.gov/CBBresearch/Yu/raid\_aps/index.html}.\\
Relevant binaries for Linux, Windows, and Mac OS X are available from the same page. 
}


\section*{Introduction}
Gaining popularity in 
 biology over the last decade, mass spectrometry (MS) has become the core 
 technology in the field of proteomics. Although this 
 technology holds the promise to identity and quantify
 proteins in complex biological mixtures/samples, such a goal 
 has not yet been achieved due to the presence of a number
 of difficulties ranging from experimental design and experimental protocol standardization to 
 data analysis~\cite{ABJH_etal_2007,FWSA_etal_2007,LO_2009}. This paper mainly focuses on
  the data analysis, especially providing accurate statistical
   significance assignments for peptide candidates in peptide identifications.
 \rev{There are many peptide identification methods that are
 available to the proteomics community. 
Because different
identification methods process (filter) the MS/MS spectra differently and
also have different scoring functions, it is natural for users to wish to compare 
search results from different search methods or to combine these results 
 to enhance identification confidence. Nevertheless, 
 there are important issues to be addressed prior to successfully reaching this goal.}

\rev{Due to intrinsic experimental variability, 
differences in the peptide chemistry, peptide-peptide interactions, ionization sources, and 
 mass analyzers used, it is natural to expect among tandem mass spectra variations in signal to noise ratios 
 even when each peptide in the mixture has equal molar concentration. 
 That said, one anticipates the noise in a mass spectrum to be spectrum-specific and 
  the  meaning of a search score depends on its context, {\it i.e.}, the spectrum used. 
 That is, although search score can be used to compare candidate peptides associated with the same
  query spectrum, it is no longer a valid measure when one wishes to compare peptides identified 
  across spectra. Not only posing a challenge for ranking identified peptides within a single experiment,
   this also raise a serious problem when one wishes to compare or combine search results from
    different scoring functions (or search methods).  
}

\rev{If one knows how to {\it translate} the score or reported $E$-value of one method 
to that of another method, or to a universal standard, it helps significantly the task of comparing/combining 
 search results. This is particularly true when} one wishes to combine search results from multiple scoring functions. 
 We showed in an earlier publication~\cite{GYWGetal_2007} that it is possible to use
  the textbook-defined $E$-value as that universal standard. 
 Providing an $E$-value calibration protocol, we demonstrated the   
  feasibility 
 of translating either the score or heuristic $E$-value reported by any method to the 
 textbook-defined $E$-value, the proposed universal statistical standard. 
 This protocol, although robust, may (a) lose spectrum-specific statistics, and may 
 (b) require a new calibration when changes in experimental set up occur. 

\rev{
Without attempting a universal statistical standard, several machine-learning based 
approaches have been developed to either re-rank identified candidate peptides~\cite{KNKA_2002,KCWNetall_2007A} 
 or to combine search results from several search methods~\cite{STN_2008A,EWT_2009A}.  
These approaches require for their analyses training data set(s), either pre-constructed or obtained on-the-fly,
 to aid the parameter selections for their discriminant functions. 
 For methods with feature vector (allowed to contain some spectrum-specific quantities) 
 updated on-the-fly~\cite{KCWNetall_2007A,EWT_2009A}, 
  the spectrum-specific bias may be partially compensated, but not giving rise to spectrum-specific statistics. 
  This is because the feature vector, although may be trained with spectrum-specific quantities,
   aims to categorize the whole training set into finite number of classes but does 
   not solely reflect the properties of any individual spectrum.  
}

 To address \rev{the issue of spectrum-specific statistics}, we developed a new MS/MS search tool, 
 RAId\_aPS (a new module of the RAId suite), that is able to  
 provide {\it spectrum-specific} $E$-values \rev{for additive scoring functions that
  do not have known theoretical score distributions}.  
  RAId\_aPS provides the users with four different modes to choose from: 
 (i) compute the total number of possible peptides (TNPP), (ii) generate score histogram,
 (iii) reassign $E$-values, and (iv) database search.  In modes (iii) and (iv), 
 RAId\_aPS is also capable of combining results~\cite{GWGFK_2008}  from different scoring functions. 
 Founded on the algorithm published earlier~\cite{GK_2008},
 mode (i) is a straight implementation of an existing idea. However, 
  modes (ii) to (iv) are novel, albeit at different levels. Mode (ii) uses the 
  algorithm published earlier~\cite{GK_2008}, nevertheless,  
  generating the {\it all-possible}-peptide (APP) score histograms of different scoring functions was never done. 
  Mode (iii) is novel from the concept to its implementation. 
    Modes (i-iii) do not have counter-parts in other components of RAId suite.    
 Mode (iv) is similar to RAId\_DbS~\cite{RAId_DbS} in the sense that it performs database searches. 
 However, the difference between mode (iv) of RAId\_aPS and RAId\_DbS lies in the use
    of statistics. The theoretical score distribution of RAId\_DbS fits score histogram of 
   database peptides per spectrum, while mode (iv) RAId\_aPS uses score distributions 
   of APP and is able to provide statistics for multiple scoring functions.

The term ``all possible peptides" (or APP) deserves some deliberation. 
 The pool of APP includes any linear arrangement of amino acids. 
 Therefore, when considering peptides of $L$ amino acids without modification,
  the APP pool includes all the $(20)^L$ combinations.  
  For the purpose of mass spectrometry data analysis, instead of peptides with a fixed length 
  one is more interested in APP within a specified molecular mass range. The number of 
  possible peptides (PP) within a molecular mass range is much larger than the
    number of database peptides within the same molecular mass range. 
 For example, for the molecular mass range $[2208{\rm Da}, 2304{\rm Da}]$, there are approximately $10,000$ peptides
  in the {\it Bos Taurus} database, while there are in total $1.385 \times 10^{26}$ PP with lengths (number of 
   amino acids) ranging from  $13$ to  $39$.

  Using dynamic programming, RAId\_aPS generates the  
   score histograms from scoring APP. These score histograms are then used to assign 
   accurate, spectrum-specific $E$-values. 
  Since RAId\_aPS uses the score histograms, or the (weighted) rank of each candidate peptide 
  considered among APP, it is already in conformity to the textbook defined $P$-value and thus 
  there is no need to {\it translate} the score or heuristic $E$-value into the universal standard. 
   Consequently, RAId\_aPS is able to provide a calibration-free 
 protocol for accurate significance assignment and for combining search results.

In order to provide a clear exposition, it is necessary for us to go into some
 technical details. Readers not interested in the details, however, may want to read
  the results section first and then come back to read other sections.  
 To make the paper easier to read and more modular, we 
 outline below the organization of this paper. 
 In the Technical Background section, we will review the similarities and differences between
  two major approaches in dealing with peptide identification statistics, describe how one may
   achieve calibration-free, spectrum-specific statistics.
  In the Method section, we first describe the dynamic programming algorithm needed to generate
   the score distribution of APP, followed by spectral
  filtering procedures each associated with a scoring function implemented.
 The incorporation of the four scoring functions are then reported since some of them are nontrivial to encode  
  via dynamic programming. We then describe how the APP statistics are
   implemented in practice, how to include modified amino acids in APP statistics, and how to combine 
  search results from different scoring functions. In the Results section,
 we describe several tests performed using various modes of RAId\_aPS,  as well as 
 the $E$-value accuracy assessment. The paper is then concluded by the Discussion section.  
  All the technical aspects that are not most essential in understanding the basic idea
 are provided either as supplementary texts or supplementary figures.    
 The most important message is that RAId\_aPS serves 
  as a calibration-free, statistically sound method for comparing or combining 
  search results from different scoring functions.

\section*{Technical Background}

Since this paper is focused on the statistical aspect of peptide
  identifications, we will start with such an example. 
 In general, it is rather easy to rank candidate  
 peptides given a tandem mass spectrum. Once a scoring function
 is selected to score peptides, 
 {\it qualified} database peptides (those within a molecular mass range and with correct enzymatic cleavages) 
 can be ranked based on their scores.
However, it becomes difficult to rank candidate peptides across all spectra.
Although a number of publications have proposed different ways tailored to deal with various aspects 
of this difficulty~\cite{GYWGetal_2007,M_2008}, this problem remains very  
 challenging. Should one take the best candidate
 peptide per spectrum and then postprocess to globally re-rank those best hits or
 should one devise something different to achieve the maximum robustness?  
 Instead of discussing the differences between these two possibilities, 
 we first wish to point out a common theme that is often unnoticed: spectrum-specificity. 

\subsection*{Spectrum Specificity}
\rev{As mentioned in the Introduction section, spectrum-specificity has not been emphasized enough. 
 However, there does exist evidence of community's recognition of this point. For example,} 
 by picking the best hit out of each spectrum, one is acknowledging spectrum-specificity,      
  because one has chosen to keep the best candidate per spectrum
 regardless of the fact that the best hit in one spectrum might have lower score
 than the second best hit in some other spectrum. In other words, by picking only the best hits 
 one has endorsed the view that the score should not be used as an objective measure 
 of identification confidence across {\it all} candidate peptides; or more precisely, 
 the meaning of score depends on its context, {\it i.e.}, the spectrum used.

There exists another route to apply the concept
 of spectrum-specificity. That is to use a spectrum-specific score distribution to
 assign an $E$-value to each candidate peptide of a spectrum.  
 Although the term spectrum-specific statistics was not explicitly mentioned,  
  the proposal of Fenyo and Beavis~\cite{FB_2003} to fit per spectrum the tail of score 
 distribution to an exponential represents the first attempt, to the best of our knowledge, in this direction. 
 The concept of spectrum-specific statistics was formally introduced by Alves and Yu~\cite{RAId}.
 The same group also developed RAId\_DbS~\cite{RAId_DbS}, so far the only database search tool 
 with a theoretically derived spectrum-specific score distribution. The importance of 
 spectrum-specific statistics is then emphasized through 
 a series of publications~\cite{GYWGetal_2007,GWGFK_2008,RAId_DbS,PGK_2005}. The key point of this type of approach 
 is to exemplify spectrum-specificity via spectrum-specific score statistics. 
After describing the common theme,
 spectrum specificity, we now turn to features associated with different types of approaches 
 to elucidate the usefulness of an even more general statistical framework.

\subsection*{Best hit per spectrum versus Accurate $E$-value}
When keeping only the best hit per spectrum, a global re-ranking among those best hits
 becomes necessary in order to decide which best hits to trust over the others. 
 This is usually achieved in one of the two ways to be described.
 The first possible choice is to use the original score 
 in conjunction with either false discovery rate (FDR) or $q$-value analysis through introduction of a decoy database.  The second choice is to use some kind of {\it refined} score in 
 conjunction with an empirical expectation-maximization-based  Bayesian approach~\cite{KNKA_2002}.   
This global re-ranking type of strategies, unfortunately, makes assumptions contradicting spectrum-specificity, a fundamental fact that is respected when only the best hit per spectrum is retained. 

 In the FDR  (be it global or local) or $q$-value analyses, 
 one pools together the best hits across spectra and order the hits by their scores. 
 This contradicts the idea of picking best hit per spectrum, 
  which essentially endorses the notion that the meaning of a peptide score is spectrum-dependent and 
 can't be used to rank peptides globally across spectra. For the Bayesian type of analyses~\cite{KNKA_2002},
 one assumes the existence of two score distributions: one for the score of  
 correctly identified spectra, in terms of best hit, and another for the score of incorrectly identified spectra. 
 This means that all correctly identified spectra --in terms of best hit-- should be ranked 
 according to  the best hit's refined  score, implying that one may use the refined score 
 to assign relative identification confidence across spectra. This again contradicts the idea 
 that the meaning of a peptide score is spectrum-dependent. Furthermore,  to perform the 
 expectation maximization procedure, one often needs to
 {\it assume} the parametric forms of the two distribution functions, which might not 
 be applicable to all scoring functions.

When the reported  spectrum-specific $E$-value (assigned to each of the candidate peptides per spectrum)  
 is in agreement with its definition, it can serve as an objective measure of identification confidence. 
 For a given spectrum 
 and a score threshold, the $E$-value associated with that score threshold is defined to be the expected 
number of false hits that have score better than or equal to that threshold. In simple
 terms, the $E$-value associated with a candidate peptide \rev{in the database} 
 may be viewed as the number of 
 false \rev{positive} hits 
 anticipated, from querying a spectrum, before calling the peptide at hand a true \rev{positive} hit.
 However, a previous study~\cite{RAId_DbS} showed that
  most $E$-value reporting methods investigated report inaccurate $E$-values.  
 To rectify this problem, we provided a protocol~\cite{GYWGetal_2007}
 to {\it calibrate} $E$-values reported by other search methods, 
 including search tools that don't report $E$-values 
 such as ProbID~\cite{ProbID} and SEQUEST~\cite{SEQUEST}. 
  However,  the calibration procedure  
 cannot  restore/recreate spectrum-specificity
 for  methods not reporting $E$-values or reporting $E$-values that are not obtained via
  characterizing the score histogram for each spectrum (spectrum-specific score modelling).
 
 Nevertheless, spectrum-specific statistics can be obtained provided that
 one extracts statistical significance from the score histogram for each   
  spectrum~\cite{GYWGetal_2007}.  
 A recent reimplementation~\cite{AYS_2009,KBJJ_2008,YALJetal_2008} of the SEQUEST XCorr follows exactly this idea. 
 To avoid possible confusion, however, we must first note that the $p^*$-value in 
 reference~\cite{AYS_2009} is actually the $E$-value. Authors of reference~\cite{AYS_2009} {\it assume} 
 that the XCorr from every spectrum can be fitted by a stretched exponential without
 providing, like most other methods, a measure on the agreement between the best fitted parametric form and
 the score distribution per spectrum. To ensure the accuracy of statistics, a measure of
 the goodness of the model~\cite{Numerical_Recipe,RAId_DbS} is actually necessary even 
 for scoring systems that have a theoretically characterized distribution.
 This is because very biased sampling might lead to a discrepancy between
 the theoretical distribution and the score distribution, not to mention a discrepancy
 between a fitted parametric form and the score distribution.

One way to circumvent the aforementioned problem is 
 to apply a target-decoy strategy at the {\it per spectrum} level. This means that one
 uses the hits from decoy database to estimate the identification confidence of peptides 
 from the target database. This approach, unfortunately, is not computationally efficient 
 because one will need a decoy database that is much larger than the target database 
 in order to have a good estimate of the $E$-value for each hit in the target database. 
  For example, if the number of qualified peptides in the decoy database 
  is $1,000$ times that in the target database,
  and if a peptide in the target database scores between the third and the fourth decoy hits,  
  then that peptide will acquire an $E$-value between $3\times 10^{-3}$ and $4\times 10^{-3}$.
  And if there are target hits that score better than the best decoy hit, all one can say is that
 they all have $E$-values smaller than $10^{-3}$. If one keeps increasing the size of the decoy database,
  one will eventually be able to {\it globally} rank the candidate peptides
 from all spectra using $E$-value. However, computational efficiency prevents us from using this strategy. 

 These aforementioned problems associated with obtaining spectrum-specific statistics can be avoided provided that 
 one uses a search method that has a theoretically derived score distribution~\cite{RAId_DbS}.
 However, restricting to methods that have theoretically derived statistics is not necessarily the
 best strategy since each search method does have different strengths~\cite{GWGFK_2008,CMI_2008}. 
 It can be advantageous to combine different types of search scores. 
 Therefore, for assigning peptides' identification confidence, it is desirable to have
 a unified framework which we now turn to.  

\subsection*{APP Statistics (calibration-free)} 
\label{sec:denovo_stat}
Alves and Yu in 2005 proposed~\cite{RAId} using the {\it de novo} rank as the statistical significance measure. 
 Despite the simplicity of this idea, it was never fully carried out. Since 
  it is this idea that inspired the development of RAId\_aPS, we need to describe the 
  basic concept to some detail so that various extensions employed in RAId\_aPS can be properly
   explained.  

 The fundamental idea is as follows. For a given MS/MS spectrum $\sigma$ with parent molecular mass $MW$
 and a given mass error tolerance $\delta$, we denote by $\Pi (\sigma, \delta)$ 
 the set of APP subjected to 
 enzymatic cleavage condition in the mass range $[MW-\delta, MW+\delta]$. 
 We also denote by $\Delta (\sigma, \delta, C)$ the set of peptides in the (target) database, subjected to 
 a set of conditions $C$, in the mass range $[MW-\delta, MW+\delta]$. The set of conditions $C$ may contain,
  for example, the enzymatic cleavage constraints, number of miscleavage sites per peptide allowed,
   and others~\cite{GYK_2008}. The following argument is also applicable to the case when one wishes to 
 weight each peptide in the APP set by its elemental composition. This may be used to form a  
 background model mimicking the amino acid composition in the target database~\cite{GK_2008,SNA_2008}.

 Let $N(S,\sigma)$ be the (weighted) number of peptides out of $\Pi (\sigma, \delta)$ 
 that have scores greater than 
 or equal to $S$. We then define the APP $P$-value corresponding to score $S$ 
 by $N(S,\sigma)/|\Pi(\sigma, \delta)|$, with $|\Pi(\sigma, \delta)|$ 
 representing the total (weighted) number of peptides 
 in the set $\Pi(\sigma, \delta)$. In general, for a given spectrum $\sigma$ 
 and a score cutoff $S$, the $P$-value $P(S\vert \sigma)$ refers to
 the probability for a {\it qualified} random peptide 
to attain a score greater than or equal to $S$ when using spectrum $\sigma$ as a query.  
 If a database contains $N_d$ qualified, unrelated  
 random peptides, one will expect to have 
 $E(S|\sigma) = N_d P(S\vert \sigma)$ number of random peptides to have quality score greater than 
 or equal to $S$. This expectation value $E(S|\sigma)$ is by definition 
 the $E$-value associated with score cutoff $S$. 

The $E$-value associated with a peptide of score $S$ using the APP $P$-value 
will therefore be 
\[
E(S\vert\sigma) = |\Delta (\sigma, \delta, C)| \frac{N(S,\sigma)}{|\Pi(\sigma, \delta)|}
\]
 where the spectrum-specific $E(S\vert \sigma)$ represents the $E$-value for a hit with score $S$ 
when the spectrum $\sigma$ is used as the query and $|\Delta (\sigma, \delta, C)|$ represents the total
 number of peptides in the set $\Delta (\sigma, \delta, C)$.  When cast in the aspect of per spectrum 
 target-decoy approach, $\Pi(\sigma, \delta)\setminus \Delta (\sigma, \delta, C)$ represents the 
 largest possible decoy database, which is supposed to provide numerically the finest $E$-values for 
 candidate peptides in the target database. (The symbol $\setminus$ is called ``setminus".  $A\setminus B$ can  be 
  called $A$ minus $B$ in the set sense or called complement of $B$ provided that set $A$ is the largest set considered and 
   every set is a subset of $A$.) Let $N'(S \vert \sigma)$ be the (weighted) number of peptide hits in
 the target database with score greater than $S$. The per spectrum target-decoy approach will
  have
\[
E(S \vert \sigma) = |\Delta (\sigma, \delta, C)| \frac{N(S,\sigma) - N'(S,\sigma)}{|\Pi(\sigma, \delta) \setminus 
 \Delta (\sigma, \delta, C)|}
 \approx |\Delta (\sigma, \delta, C)| \frac{N(S,\sigma)}{|\Pi(\sigma, \delta)|}
\]  
 where the last result comes from $N'(S,\sigma) \ll N(S,\sigma)$ and 
 $|\Pi(\sigma, \delta) \setminus \Delta (\sigma, \delta, C)| \approx |\Pi(\sigma, \delta)|$ for any practical applications.

 For a typical molecular mass of $1500$ Dalton (Da) and in the absence of weighting, 
 $|\Pi(\sigma, \pm 1Da)| \approx 5\times 10^{15}$. For a typical organismal database, 
 such as that of {\it Homo sapiens}, the total number of peptides within the molecular mass range
  without any condition is only  $|\Delta (\sigma, \pm 1Da)| \approx 3\times 10^{3}$. 
 Therefore, $  5\times 10^{15}  \ge |\Pi(\sigma, \pm 1Da) \setminus \Delta (\sigma, \pm 1Da, C)| \ge  5\times 10^{15} - 3\times 10^{3}$, and $|\Pi(\sigma, \pm 1Da) \setminus \Delta (\sigma, \pm 1Da, C)| \approx  5\times 10^{15} $. 
In the presence of peptide weighting, one still has $|\Pi(\sigma, \pm 1Da)|/|\Pi(\sigma, \pm 1Da) \setminus \Delta (\sigma, \pm 1Da,C)| \approx 1$.  Therefore, $|\Pi(\sigma, \delta) \setminus \Delta (\sigma, \delta, C)| \approx |\Pi(\sigma, \delta)|$. As for $N'(S,\sigma)$ versus
 $N(S,\sigma)$,  by definition $N' = 0$ for best target hit and $N(S,\sigma)$
  typically increases much faster than $N'(S,\sigma)$  when $S$ is lowered, 
 thus $N'(S,\sigma) \ll N(S,\sigma)$, a fact also observed in reference~\cite{SNA_2008}.
 Consequently, $N(S,\sigma)-N'(S,\sigma) \approx N(S,\sigma)$ is a very good approximation. Therefore, 
 the APP statistics also serve as the best per spectrum target-decoy statistics. 
 The only question now is how does one get the score distribution of APP? 

It turns out that if the score of a peptide is the sum of {\it local} contributions, meaning each term 
 in the sum is uniquely determined by specifying a fragment's m/z value, then it is possible to
 construct the score histogram of APP via dynamic programming~\cite{GK_2008,SNA_2008}. 
  When there exists intrinsically nonlocal contribution in peptide scoring, 
 it is no longer possible to obtain the 
 full histogram by dynamic programming. However, it is still possible to estimate the {\it de novo} rank via
  a scaling approach~\cite{PGK_2005} similar to that used in statistical physics. 
 The key point, as will be shown later, is that for the four scoring functions implemented 
  in RAId\_aPS,  by using the APP statistics, it is no longer critical to theoretically 
  characterize the  score distribution obtained from the database search. 
  This is because the $E$-value obtained via RAId\_aPS does agree 
  well with the textbook definition. The APP statistics employed 
 by RAId\_aPS may be extended to provide robust spectrum-specific statistics for scoring functions 
 that do not have theoretically characterized score distributions.  
One advantage to having a method that can provide robust spectrum-specific statistics for 
 different scoring functions is that if 
  the $E$-value reported by each method agrees with its definition, 
  one can {\it compare} and {\it combine} search results from different search methods~\cite{GWGFK_2008}.

\section*{Methods}

\subsection*{Basic Dynamic Programming Algorithm}
To generate the score histogram of APP in a speedy manner, 
RAId\_aPS does not score every  possible peptide individually. As a matter of fact,
 it is impossible to score every  possible peptide individually. For example, consider 
 a typical parent ion molecular mass of $1,500$ Da. It can be shown that the TNPP within $1$ Da of 
 this molecular mass is more than $10^{15}$. Even if 
 one has a simple scoring function and a fast computer that can score one hundred millions 
 peptides per second, it will take more than $116$ days of 
 computer time to generate the score histogram for a single spectrum. 
 
 In real application, one needs to
 analyze a spectrum in a short time.  How could one achieve this? 
 One may use a  1-dimensional (1D) mass grid to encode/score 
 APP~\cite{GK_2008,SNA_2008}.   
At each mass index of the grid, the local score contribution associated with all partial
 peptides reaching that location is computed only once and this information may be 
 propagated forward to other mass entries via dynamic programming, making it possible to generate 
 the score histogram of APP without individually scoring all peptides. 
 In the score histogram, instead of counting number of peptides associated with a certain score, 
 it is also possible to weight each peptide sequence according to its elemental composition.
 For a peptide sequence $[a_1,a_2,\ldots,a_M]$, one may assign it a weight~\cite{GK_2008,SNA_2008} 
 $p(a_1) p(a_2)\ldots p(a_M)$ with $p(a_i)$ being the emitting probability of amino acid $a_i$.
 
\begin{figure}[htbp]
\begin{center}
\includegraphics[width=0.95\textwidth,angle=0 ]{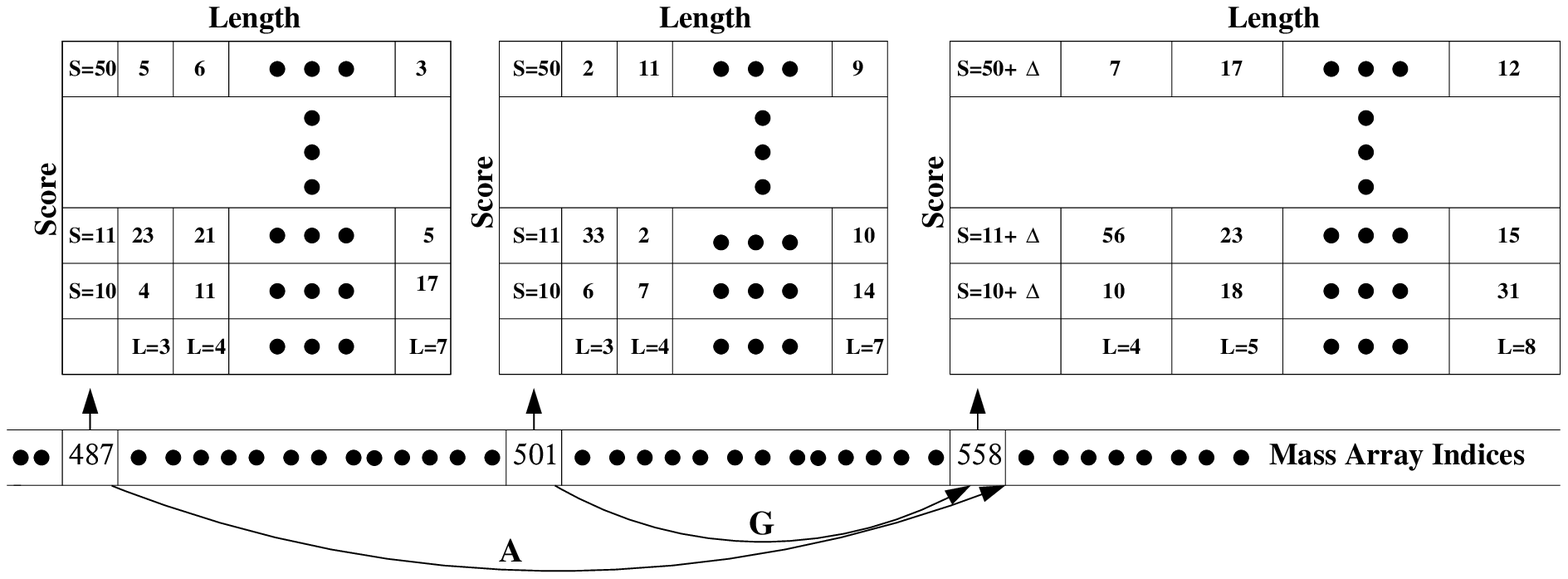}
\end{center}
\caption{\small Illustration of APP mass grid with internal structure. 
In addition to show the basic mass grid, 
 this figure illustrates,using the peptide lengths as an example,  
 the possibility of including additional structures
  in the (raw) score histogram associated with each mass index. The basic idea 
 of obtaining the score histogram via dynamic programming is explained in
  the Method section. The key step to incorporate additional structure
  is to let the (weighted) count associated with each (raw) score be further
  categorized by the lengths of partial peptides reaching each mass index. 
  In the end, one will apply the 
 length correction factor to the raw score to obtain the real score histogram.  
 Apparently, one may also keep track of the number of $b$ ($y$) peaks accumulated 
 within the raw score histogram. Again, the factorial contribution can be added at
 the end prior to the construction of the final score histogram.
}
\label{DP.grid.2}
\end{figure}

	For illustration purposes, the mass grid of 1Da resolution is used in Figure~\ref{DP.grid.2}.  
Each mass index contains a score histogram, with each entry in the left column indicating a score 
and the corresponding entry at the right column recording the number of partial peptides reaching that mass index with
   that score. The score histogram is obtained using a backtracking update rule. 
 For example, at the mass grid $558$, the local score contribution from evidence peaks 
 in the spectrum is assumed to contribute $\Delta$ amount of score. Looking back to mass grid
 $501$ ($57$ Da less than $558$ Da), one knows that by attaching a glycine residue 
  to the partial peptides reaching mass index $501$ one will then advance 
 these peptides to index $558$. 
 Similarly, any partial peptides reaching mass index $487$ will move to mass index $558$ by
 adding an alanine residue. Therefore, at mass index $558$ the score histogram is the 
 superposition of score histograms associated with the other twenty lighter mass grids corresponding 
 respectively to the twenty amino acids. For simplicity, the illustration is drawn as if there are 
 only two amino acids, glycine and alanine. When one weights each peptide by its elemental 
 composition, the counts next to the scores in the histogram are weighted and no longer 
 integers. For example, the weighted count $n(558)$ at mass index $558$ will be 
 given by $n(558) = \sum_{a=1}^{20} p_a\; n(558-m_a) $ where $m_a$ is the
 mass of amino acid $a$ rounded to the nearest Da and $p_a$ is the emitting probability associated with 
 amino acid $a$. In addition to attaching a score histogram to each mass grid, one may also include 
  other internal structures such as peptide lengths, peak counts, etc. as shown in the caption 
   of Figure~\ref{DP.grid.2}.  
 When one suppresses the score and only counts number of partial
 peptides reaching a certain mass index, the update rule readily provides the total number
 of peptides within a given mass range.

\subsection*{Spectral Filtering}
Before describing the scoring functions, the major component of peptide database search tools,
 we first mention spectral filtering, an often under-emphasized but equally important ingredient. 
 Starting with a raw tandem mass spectrum, spectral filtering produces a processed spectrum
 that is used to score candidate peptides in the database. 
 Apparently, information kept in the processed spectrum plays an important role
 in the effectiveness of a tool's performance in database searches.
  Customized for different scoring functions, different filtering strategies 
 are employed by different search tools. In order for RAId\_aPS to 
  capture the essence of a scoring function, it is very important for RAId\_aPS
  to produce, for every input raw spectrum, a filtered spectrum that is as close as possible to 
  the one produced by other search tool's filtering protocol. For most 
  search tools, the filtering heuristics are not clearly documented. 
 For that reason, it becomes necessary to delve into the source code of the search 
 program to find out each method's spectral filtering protocol. We are thus 
 limited to search tools whose source programs are available or those with filtering strategies clearly documented.

 For RAId score, the spectral filtering strategy was described in  
 an earlier publication~\cite{RAId_DbS}. 
For Hyperscore~\cite{XTandem}, XCorr~\cite{SEQUEST}, and K-score~\cite{BKCM_2006,AJNJ_etall_2005}, 
  the details of spectral filtering will be described in Text S1. 
 Since the SEQUEST source code is not available, for XCorr score we attempt to replicate
 the filtering of Crux~\cite{YALJetal_2008}, a search method that has been shown to reproduce SEQUEST XCorr~\cite{YALJetal_2008}. 
 That the filtering strategies extracted are accurate can be seen from 
  Figure~S1. The spectral correlation histograms
  between the filtered spectra produced by RAId\_aPS's Hyperscore/XCorr/K-score with the filtered 
 spectra from X!Tandem/Crux/X!Tandem(with K-score plug-in) show that   
  RAId\_aPS is able to produce 
 filtered spectra identical to those generated by the canonical programs.  
  Although the spectral filtering strategies associated with various search tools investigated seem
  stable, it is still possible that the developers may change their filtering strategies in the future. 
   When that happens, one should be able to update RAId\_aPS to reflect 
  the filtering changes provided that the source programs are still accessible and
 clearly documented.

\begin{figure}[htbp]
\begin{center}
\includegraphics[width=0.95\textwidth,angle=0 ]{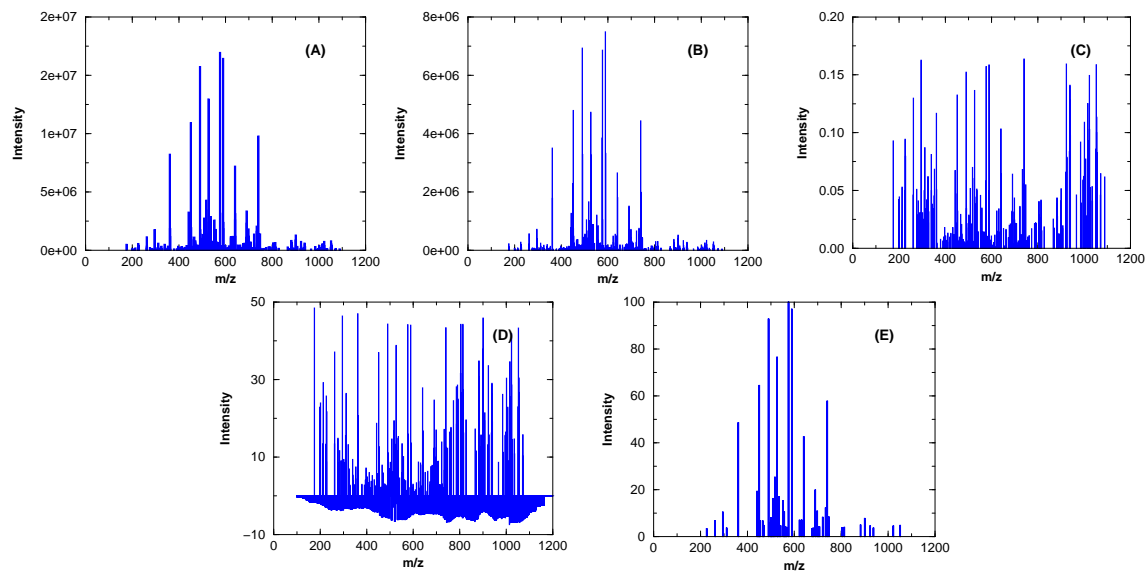}
\end{center}
\caption{\small  Example processed spectra from different scoring functions versus the 
 original spectrum.  The centroid spectrum used has a parent ion mass of $1640.80$ Da. 
In panel (A),  the  original spectrum is displayed;
 (B) shows the processed spectrum generated by the filtering protocol of RAId\_DbS scoring function; 
 (C) exhibits the processed spectrum generated by the filtering protocol of K-score; while
  (D) and (E) correspond respectively to the processed spectra produced by 
      XCorr and Hyperscore.
}
\label{Spect_filter}
\end{figure}

\begin{figure*}[htbp]
\begin{center}
\includegraphics[width=0.95\textwidth,angle=0]{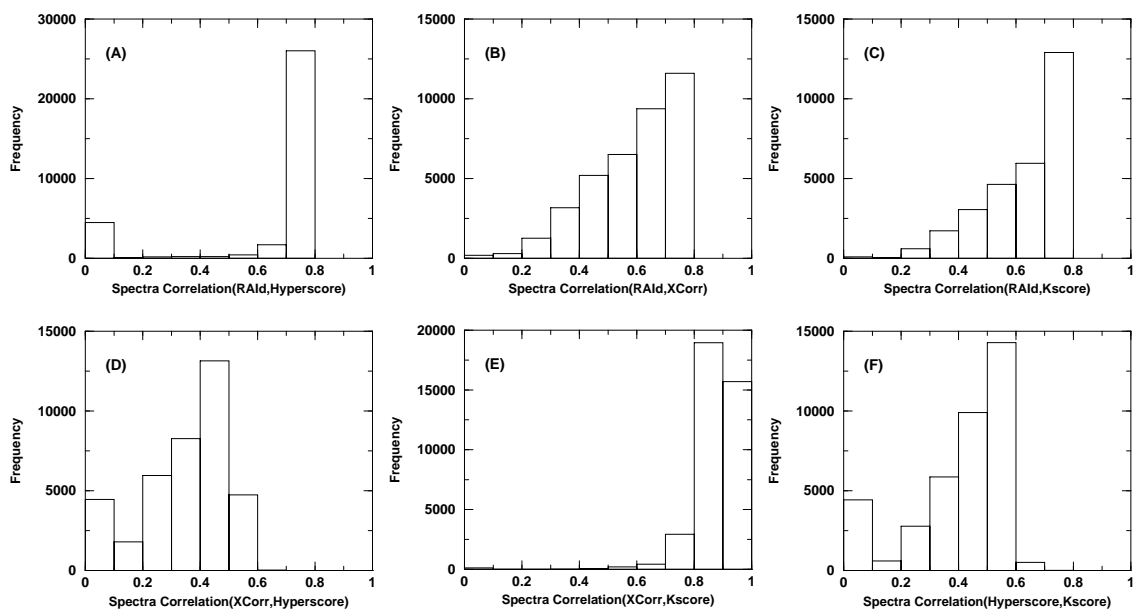}
\end{center}
\caption{ Histograms of correlations  between filtering strategies. 
  Used in this plot are $38,424$ raw centroid spectra from the ISB data set~\cite{KSNSetal_2002}. 
 Each raw spectrum
 will have four different processed spectra come from each of  the four different filtering strategies. 
 The mass fragments of every filtered spectrum 
 are then read to a mass grid. The spectrum is then viewed as a vector 
 with non-vanishing components only at the  populated component/mass indices. One then 
 normalizes each {\it filtered} spectrum vector to unit length. An inner product of any two
 filtered spectral vectors represents the correlation between them. When the spectral 
 quality does not pass a method-dependent threshold, the corresponding  
 filtering protocol may turn the raw spectrum into a null spectrum without further 
 searching the database. For a given pair of filtering methods and a raw spectrum, if each of
 the two filtering methods produces a nonempty filtered spectrum, one may turn those
 filtered spectra into spectral vectors and compute their inner product, i.e., their
 correlation. For each pair of filtering methods, these inner products are accumulated
  and plotted as a correlation histogram. All six pairwise combinations are shown. 
} \label{spect_corr_centroid}
\end{figure*}

 Instead of elaborating on various filtering strategies, 
 let us first use a experimentally obtained spectrum to demonstrate the effect of spectral 
 filtering employed by different methods. Figure~\ref{Spect_filter} shows the raw spectrum,
 and the filtered spectra processed by the four scoring methods mentioned. 
 The general trend is as follows: RAId score usually produces 
 the filtered spectrum that resembles the original spectrum the most; 
 Hyperscore filtering also produces 
 a processed spectrum that is similar to the original spectrum; for XCorr and 
 K-score  the filtered spectra in general look quite different from the original spectrum. 
 The differences in the filtered spectra might be a major factor contributing to the 
 fact that different search methods have different and often complementary strengths.  
  The correlation between any pair of filtering strategies can be quantified. 
 Starting with a large set of raw spectra, one may 
 process these spectra with a pair of different methods. For each raw spectrum, one obtains 
 two different filtered spectra and can compute their correlation. 
 The correlation between every pair of filtered spectra can then be collected to form the
 correlation histogram, reflecting the correlation between a pair of filtering strategies.  
 Figure~\ref{spect_corr_centroid} and Figure~S2 exhibit the correlation histograms between each pair of
 filtering strategies using different data types:  
 centroid (A1-A4 of ISB data set~\cite{KSNSetal_2002}, Figure~~\ref{spect_corr_centroid}) 
 and profile (NHLBI data set~\cite{GYWGetal_2007}, Figure~S2). 
 The large correlation between XCorr and K-score
 may be the cause of their significant scoring correlation observed.

\subsection*{Scoring Functions}

To better express the scoring functions, let us first define the following notations. 
 For a given peptide $\pi$, the set of corresponding theoretical mass over charge (m/z) ratios taken into consideration by a scoring
 function is called $T(\pi)$, which is also used to indicate the number of elements in the set $T(\pi)$
 whenever no confusion arises. The set $T(\pi)$ varies from software to software. 
 However, the fragmentation series 
 $(a_n,b_n,b_n\!\!-\!\!18, b_n\!\!-\!\!17, c_n,x_n,y_n,y_n\!\!-\!\!18, y_n\!\!-\!\!17,z_n)$
 include what most methods consider. 
 The Heaviside step function $\theta(x)$ is defined by $\theta(x<0) = 0$ and $\theta (x > 0 ) = 1$. 
 We introduce $I_i$ as a shorthand notation for $I(m_i)$, the peak intensity associated with theoretical mass $m_i$ in the {\it processed} spectrum. In an experimental spectrum, the mass 
 giving rise to $I_i$ usually does not coincide with $m_i$.  
 The absolute difference between the experimental mass (giving rise to $I_i$)
  and the theoretical mass $m_i$ is denoted by $\Delta m_i$. 
The notation $I'_i$ is used in place of $I_i$ when 
 the preprocessing of the spectrum involves a nonlinear transformation of the peak intensity or 
 involves generation of additional peaks. We now list the four different scoring function implemented:
\bea
{\rm RAId~}S(\pi) &=& \frac{1}{T(\pi)} \sum_{i=1}^{T(\pi)} \ln(I_i)\;  e^{-\Delta m_i} \theta(1-\Delta m_i) 
 \label{DbS_score} \\
{\rm Hyperscore~} S(\pi) &=& 4 \log_{10} \left[  \left( \sum_{i=1}^{T(\pi)}\; I'_i \right) b\, !\;  y! \right]
\label{Hyper_score}\\
{\rm XCorr~} S(\pi) &=&{1\over 10000} \sum_{i=1}^{T(\pi)}\; w_i I'_i
\label{XCorr_score}\\
{\rm K\!\!-\!score~} S(\pi) &=&  {1000 \, \ln (l) \over 3\sqrt{l}}
\sum_{i=1}^{T(\pi)}\;  w_i I'_i
\label{K_score}
\eea

The first scoring function listed is 
employed by RAId\_DbS~\cite{RAId_DbS};  
the second one mimicks the Hyperscore (${\mathrm X}_{II}$) 
of X!Tandem~\cite{FB_2003}; the third one mimicks the 
XCorr score used in SEQUEST and is similar to what was implemented 
 in Crux~\cite{YALJetal_2008,KBJJ_2008}; the last one mimicks K-score~\cite{BKCM_2006}, a plug-in for X!Tandem.    
 For the RAId score, the set $T(\pi)$ includes only the $b$- and $y$-series peaks. 
 For the Hyperscore, $T(\pi)$ includes $\{ b_n,~y_n  \}$. For XCorr, 
  $T(\pi)$ includes $\{b_n,~y_n,~b_n-1,~b_n+1,~y_n-1,~y_n+1,~b_n-18,~b_n-17,y_n-17,~a_n\}$ with the
 corresponding weights given by $\{ 50,~50,~25,~25,~25,~25,~10,~10,~10,~10 \}$.
 For K-score, $T(\pi)$ includes $\{b_n,~y_n,~b_n-1,~b_n+1,~y_n-1,~y_n+1 \}$ 
 with the corresponding weights given by $\{ 1,~1,~0.5,~0.5,~0.5,~0.5\}$. 
  To speed up the code, we have chosen to rescale the weights for XCorr 
  (see the ``Crux Filtering and XCorr" section of Text S1 for detail). 

Very often it is useful to include the peptide length in the scoring of a peptide. 
Using RAId score as a simple example, two peptides of length $11$ and $16$ may achieve 
the same raw score $S'_{11}=S'_{16}=10$, sum of the logarithm of evidence peak intensity. 
 A longer peptide consists of a longer list of theoretical peaks to look for
 and may thus score higher by chance. RAId\_DbS scoring function~\cite{RAId_DbS} deals with this
 issue by dividing the raw score by the length of the 
 theoretical peak list. Upon doing so, 
 one has $S_{11} = S'_{11}/(2\times(11-1))=1/2$ and  $S_{16} = S'_{16}/(2\times(16-1))=1/3$. 
 This score normalization may help in discriminating true positives from false positives.  
  The other scoring function utilizing 
 the peptide length information is the K-score. Hyperscore, employed  by X!Tandem, 
  uses a slightly different score renormalization strategy. 
  Inside the logarithm, the Hyperscore  contains two factorials, $b!$ and $y!$.
  For each candidate peptide, $b$ ($y$) 
  represents the total number of $b$-series ($y$-series) evidence peaks found in the spectrum. 
 At any specified mass index in the mass grid, unlike the peak intensity associated with that index,   
 neither the peptide length nor the total number of the b (y) peaks has a unique corresponding value. 
 Therefore, one needs to extend the basic algorithm outlined in the previous subsection to accommodate
 these additional information needed for scoring.

As documented in reference~\cite{GK_2008}, it is possible to introduce additional
 structures in the score histogram associated with each mass index. 
 The flexibility to introduce additional structures of various dimensions makes 
RAId\_aPS a  versatile tool: it can accommodate 
the scoring functions that utilize length information or the number 
of $b$-series ($y$-series) peaks to compute the final peptide score. 
  Using peptide length as an example, Figure~\ref{DP.grid.2} demonstrates the inclusion of additional structures. 
 More detailed exposition about the inclusion of internal structures can be found in reference~\cite{GK_2008}.

\begin{figure}[htbp]
\begin{center}
\includegraphics[width=0.95\textwidth,angle=0 ]{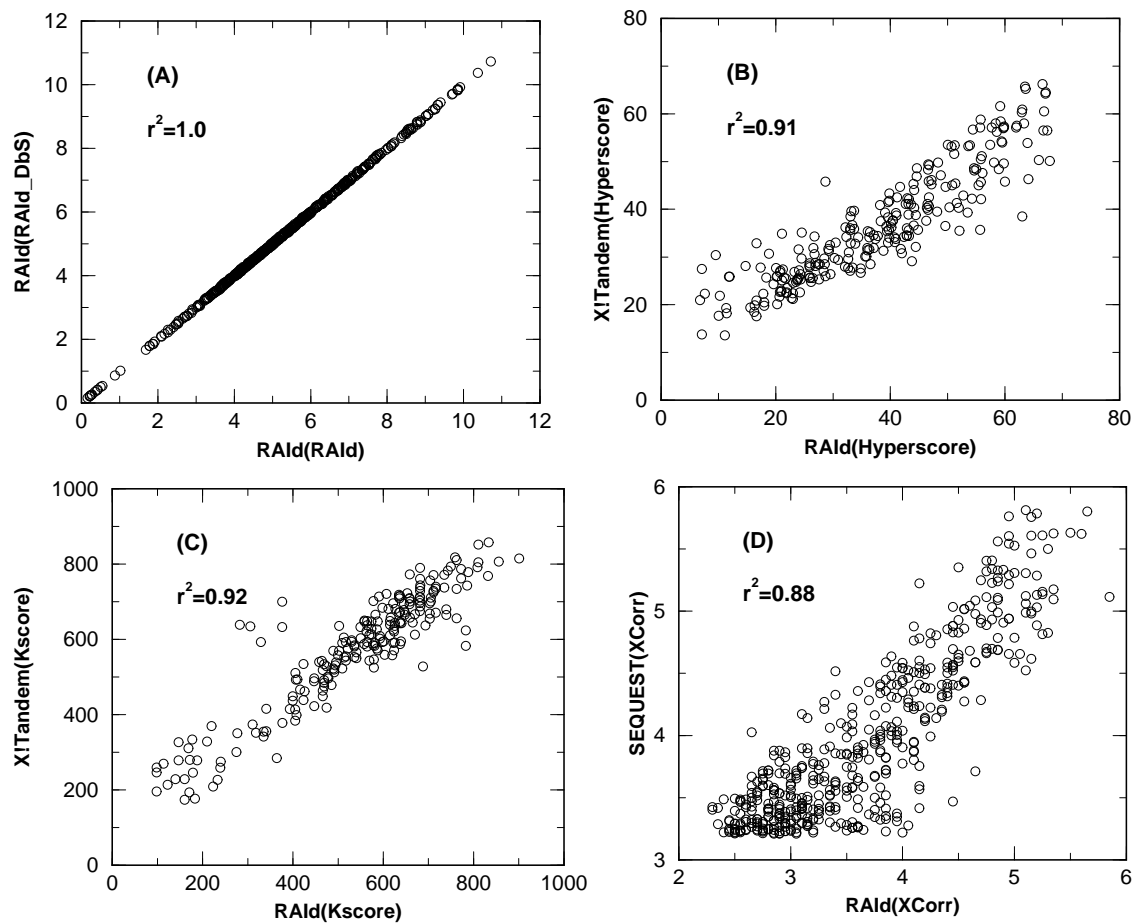}
\end{center}
\caption{\small Score correlations. A subset of the ISB centroid data set~\cite{KSNSetal_2002} was used to perform this evaluation.
 For each scoring function, when the best hit per spectrum (analyzed using the analysis program that the scoring function was originally used for) is a true positive, that candidate peptide is scored again using the 
 corresponding scoring function implemented in RAId\_aPS. Each true positive best hit thus gives rise to
 two scores and plotted using the following rule:  the first score is used as the ordinate while
 the second score (from RAId\_aPS) is used as the abscissa. 
   Including $500$ spectra, panel A is for the RAId score.  
 Panel B is for Hyperscore and contains $248$ spectra. The result of K-score is
 shown in panel C with $220$ spectra.  Shown with $500$ spectra, panel D documents the results for XCorr. 
}
\label{score_corr_centroid}
\end{figure}

Although the spectral filtering parts of various scoring functions are 
 replicated exactly, a candidate peptide may receive 
 different scores from RAId\_aPS and the original programs. 
  This phenomenon can be seen in Figure~\ref{score_corr_centroid}: the ordinate of each data
   point displays the search score of the best hit of a centroid  spectrum using the original programs, 
   while the abscissa of the same data point shows the score reported by RAId\_aPS.  
   The corresponding plots for profile data are shown in Figure~S3. 

 The major source of score difference is due to RAId\_aPS's omission of
 {\it heuristics} while implementing a published scoring function. 
 For each scoring function, many scoring heuristics are present in the source
 code. While some of the heuristics cannot be included via dynamic programming, all
  these heuristics are either not described or not justified in the original papers. 
  For these reasons, RAId\_aPS does not include those unpublished heuristics. 
 Therefore, the Hyperscore/XCorr/K-score scoring functions 
 implemented in RAId\_aPS should be regarded as our attempt to mimick 
 the original Hyperscore/XCorr/K-score scoring functions. Although the scoring functions
 we implemented are not exact replicas of the original ones, due to omission of heuristics, 
 we can see from Figure~\ref{score_corr_centroid} (and also Figure~S3 when
  tested on profile data) that there exist strong correlation between each scoring function implemented in
 RAId\_aPS and the original, corresponding scoring function. In other words, the scoring functions
 implemented in RAId\_aPS do capture the essence of these original scoring functions.

\subsection*{APP Statistics: practical implementation} 
In the APP statistics section, we described how to use APP statistics to obtain 
 $P$-values and $E$-values with or without weighting each peptide by its elemental composition.
 In this subsection, we will complement the theoretical presentation by describing some pragmatic aspects
 of the implementation. 
 
In order to build the score histogram quickly, it is necessary to discretize the score, thereby compromising 
 to some degree the score precision. However, this rounding of scores does not affect peptide
 scoring when using RAId\_aPS as a database search tool or a tool to provide statistical significance
 for a list of peptides.  
Specifically, the evidence score collected at each mass index is
 stored in two formats: one with much higher precision and the other rounded to nearest integer. 
 The rounded values are used in dynamic programming to propagate the score histogram forward, 
 facilitating a speedy construction of the score histogram. The slight error introduced in 
 individual peptide scoring does not influence the accuracy of the score histogram much
 since these errors largely cancel each other when lumping the scores into
 a histogram. In the database search mode, 
 RAId\_aPS will sum the high precision evidence scores in the mass indices traversed by 
  the candidate peptide being scored. Therefore the score associated with each candidate
 peptide in the database search mode has a better resolution than that in the score histogram. 
 To obtain the statistical significance associated with each candidate peptide, RAId\_aPS
 performs an interpolation procedure to obtain the $P$-value, 
 \[
 P(S,\sigma) = \frac{N(S,\sigma)}{|\Pi(\sigma, \delta)|}\;.
 \] 
 Multiplying the $P$-value by the number of qualified peptides $|\Delta (\sigma, \delta, C)| $ in the target 
database provides the $E$-value
\[
E(S,\sigma) = |\Delta (\sigma, \delta, C)|~P(S,\sigma) \; .
\] 

\subsection*{APP Statistics including PTM amino acids} 
Since proteins do contain PTM amino acids, it is important
 for peptide identification tools to consider amino acid modifications
 in the statistical analysis. By scoring only qualified peptides, 
 database search methods have little problem including PTM amino acids
 provided that the score distribution is theoretically characterizable. 
 For APP based statistics, even though the score distribution is
  not always characterizable, information from {\it qualified} peptides 
 in database search may be used to generate the emission probabilities of
  all the amino acids, PTMs included, needed for APP based statistics.    
 
  Given a parent ion mass and a database, once the allowable PTMs are specified,
    the number of peptides along with possible types of modifications are fixed. 
  This renders a parent-ion-mass specific and database specific emission probabilities 
   for PTMs. Nevertheless, the list of qualified peptides may vary with molecular mass
    error tolerance while the allowable PTMs may also vary with users' specification for
  a search. Once the list of qualified peptides for a spectrum is given, the emission probabilities  
   of each amino acid (including PTMs) are computed as follows:            
  for each amino acid $B$, RAId\_aPS first counts the number of 
 occurrences of the unmodified amino acids $n(B)$ and the number of occurrences $n(B_i)$ of
 $B$ modified into a different form $B_i$, with $i=1,\ldots,k$. 
 RAId\_aPS then proportionally distributes the emission probability
 $p_0(B)$ associated with amino acid $B$ to all the possible modified forms using the following formulas 
\bea
p(B) &=& \frac{n(B) + 1}{n(B)+1+\sum_{i=1}^k n(B_k)} \; p_0(B) \label{a.reg.bg} \\
p(B_i) &=& \frac{n(B_i)}{n(B)+1+\sum_{i=1}^k n(B_k)} \; p_0(B)  \label{a.mod.bg} \; .
\eea
Effectively, one pseudocount is always given to each unmodified amino acid. 

Therefore, for a given list of peptides, RAId\_aPS will count the total number 
 of distinct amino acids modifications. In principle, RAId\_aPS can 
 incorporate all those modified amino acids in the score histogram construction.
 However, for reasons to be described below, RAId\_aPS retains
  no more than the ten most abundant PTMs in calculating the new emission probabilities.
  First, the estimated emission probabilities of PTMs become less trustworthy when the occurrences
   of those PTMs are rare. Second, inclusion of many PTMs can slow down the process, although 
    not very much. Assume that one incorporates ${\mathcal M}$ modified amino acids in the 
  score histogram construction, the number of trace backs per mass index becomes 
  $20+{\mathcal M}$ instead of $20$. This introduces a factor of $(20+{\mathcal M})/20$ compared to
   the original construction. Further, the size of score array associated with each mass index needs to be
    larger than before and thus require more time to compound the score histogram. This approximately introduce another factor 
   of $(20+{\mathcal M})/20$ to the computation speed. Thus, introducing ${\mathcal M}$ modifications
    will introduce a multiplicative factor of $(1+\frac{\mathcal M}{20})^2$ to the computation time. 
 To ensure that the average run time does not grow more than two fold,  we set the maximum ${\mathcal M}$ allowed 
  to be ten.             
  The new set of {\it normalized} background frequencies 
 (with the most abundant PTMs included) may then be fed into RAId\_aPS
 to compute the corresponding APP score histogram. The histogram obtained
 is then used to calculate the statistical significance of each reported 
 peptide. 

  Although rare PTMs in the peptide list might be omitted in constructing the 
 APP score histogram, the impact on the statistical significance accuracy 
 is minute. For if one were to include those PTMs, due to their small 
 normalized emission probabilities, peptides containing those PTMs 
 would be weighted substantially less than others and thus  
 would not significantly affect the shape of the score histogram.  
 As for the emission probability $p_0(B)$ ---needed in eqs.~(\ref{a.reg.bg}-\ref{a.mod.bg})---
  associated with amino acid $B$, one may use either known amino acid background frequencies such as the Robinson-Robinson~\cite{RR_1991}
  frequencies or can calculate the number of occurrences of all amino acids in a {\it parent-ion-mass-specific} and 
   {\it database-specific} manner. The former approach is adopted by RAId\_aPS when the number of peptides (provided by
    the user or extracted from the database) is less than $2,000$; otherwise, the latter approach is employed.     
 There exists, of course, room for improvement in terms of including 
 PTMs in the APP statistics. Alternatives are currently under investigations.

\subsection*{Combining Search Results from Different Scoring Functions}
 When the user select multiple scoring functions in mode (iii) and mode (iv), 
 RAId\_aPS is able to combine statistical significances 
 reported by the different scoring functions. For database search (mode (iv)), the protocol to combine
 search results is identical to what was described before~\cite{GWGFK_2008}.   
 In this section, we will briefly review this method. 

For a given spectrum $\sigma$, to combine search results 
 from $m$ scoring functions (say scoring function $A_1$, $\ldots$, $A_m$),  
 we first construct a union peptide list ${\mathcal L}(\sigma) \equiv 
 {\mathcal L}_{A_1} (\sigma)\cup \ldots \cup {\mathcal L}_{A_m} (\sigma)$, 
 where ${\mathcal L}_{A_i} (\sigma)$ is the reported list of peptide hits by method 
 $A_i$ for spectrum $\sigma$. A peptide in the union list has at least one,
 and may have up to $m$ $E$-values derived from APP $P$-values, depending on how many 
 scoring functions reported that specific peptide in their candidate lists. 
 Each of the $E$-values associated with a peptide will
 be first transformed into a {\it database $P$-value}~\cite{GWGFK_2008}, 
\rev{ representing the probability of seeing at least one
 hit in a given random database with quality score
 larger than or equal to $S$}. 
 If one assumes that the occurrence of a high-scoring 
 random hit is a rare  event and thus can be modeled by 
 a Poisson process with expected
 number of occurrence $E(S\vert \sigma)$, one may 
 \rev{obtain the database $P$-value mentioned} earlier via
\be \label{E2P}
P_{\rm db}(S \vert \sigma) = 1 - e^{-E(S\vert \sigma)}\; . 
\ee   

The database $P$-value of peptide $\pi$ is set to one for methods that
 do not report $\pi$ as a candidate.  
 After this procedure, each peptide in the 
 list ${\mathcal L}(\sigma)$  has $m$ database $P$-values $(P_1,P_2,\ldots, P_m)$. 
 \rev{Assume that these $P$-values are independent}, 
 the combined $P$-value \rev{(with $\tau \equiv \prod_{i=1}^m P_i$) for peptide $\pi$} is given by~\cite{GWGFK_2008}
\be
P_{\rm comb}(\pi) =  \tau \sum_{k=0}^{m} {[\ln (1/\tau)]^k \over k! }
\label{final_p} \;  
\ee
 Once $P_{\rm comb}(\pi)$ is obtained, we may 
  invert the formula in Eq.~(\ref{E2P}) to get a combined 
 $E$-value $E_{\rm comb}$ via 
\be \label{P2E}
E_{\rm comb}(\pi) = \ln \left( {1\over 1-P_{\rm comb}(\pi)} \right) \; .
\ee
 We then use $E_{\rm comb}(\pi)$ as the final $E$-value 
 to determine the statistical significance of peptide candidate $\pi$,
 similar to what is used in reference~\cite{oNAR_2006}. 
 \rev{From a theoretical stand point, one might ask whether or not eq.~(\ref{final_p}) 
  always gives rise to a smaller combined $P$-value than any of the input $P$-values.  
 The answer is no. For example, consider $P_1 = p < 1$ and $P_2 =1$. One then
 has combined $P$-value $p[1+\ln(1/p)]$ larger than $P_1$.  
 Readers interested in more details are referred to Appendix B of reference~\cite{GWGFK_2008}.}

 \rev{The combining $P$-value strategy outlined by eqs.~(\ref{E2P}-\ref{P2E})
 is founded on the assumption that $P$-values resulting from different search scores are independent. 
 That is, the resulting significance assignment is valid only when 
  scoring functions considered are uncorrelated, or at most weakly correlated.
In our earlier investigation~\cite{GWGFK_2008},  we found that although many scoring functions
 are looking for similar scoring evidences, the pairwise correlations among scoring functions
  investigated are weak, perhaps due to different spectral filtering methods employed.
  The weak pairwise correlations among different scoring functions implies that 
   the outlined strategy above may still provide decent significance assignment. 
 How to {\it properly} take into account method correlations while combining the search results 
  is of course a very important and open problem.       
 }

 Suppose one has obtained a list of candidate peptides from
 some analysis tools that provides only crude
  statistical significance assignment or no significance assignment at all, 
 it is possible to upload this list of peptides along with the spectrum to RAId\_aPS 
 to get a reassignment of statistical significance via mode (iii) of RAId\_aPS. 
 The fundamental idea here is to first obtain the score histograms
 corresponding to the list of scoring functions selected. With the histograms 
 constructed, one can generate the $P$-values for any score specified. Therefore,
 for a chosen scoring function and a given list of peptides, 
 RAId\_aPS can provide for each peptide an APP $P$-value
 by scoring each peptide and then inferring from the normalized score histogram. 

In practical implementation, RAId\_aPS sorts the list of peptides   
 according to their molecular masses and identifies their corresponding mass
 indices on the mass grid. Using these indices as terminating points, but one
 at a time, RAId\_aPS constructs score histograms assuming that the parent ion weight
 is given by the mass indices considered. Each peptide in the list is then rescored
 using the user-selected scoring versions implemented in RAId\_aPS and the $P$-values 
corresponding to these scoring functions are obtained. 
If no further information other than
  a flat list of peptides is given, RAId\_aPS will combine these $P$-values using 
 eq.~(\ref{final_p}) and return a combined $P$-value for each peptide in the list.
When the number of qualified database peptides is known --which is the case
 if one directly uploads to RAId\_aPS any of the output files of Mascot, SEQUEST, or X!Tandem--  
RAId\_aPS will first transform the $P$-values into $E$-values and then into
 database $P$-values (eq.~(\ref{E2P})). For each peptide in the list, 
 RAId\_aPS will then combine their database $P$-values using eq.~(\ref{final_p}) and then 
 obtain the final $E$-value via eq.~(\ref{P2E}).

\section*{Results}

\subsection*{$\pmb{E}$-value Accuracy}
 In the APP statistics subsection of Technical Background, 
  it was  demonstrated that statistical significance assignment based 
 on the APP score histogram is {\it spectrum-specific}. However, one must
 verify $E$-value accuracy before claiming that accurate spectrum-specific statistics
  are achieved via APP statistics. A straightforward way to test $E$-value accuracy~\cite{RAId_DbS}
   is to compare the averaged number of false positives (the textbook definition) 
 versus reported $E$-value using a spectral dataset resulting from a known mixture.    
 To be specific, one will first eliminate true positives from a database, and then use the spectra 
 from a known mixture as queries to look for peptide hits. Since the true positives are removed 
 from the database beforehand, all the peptide hits are false positives.
  One then aggregates all the false positives together --there might be many false positives from
 one spectrum-- and then sorts them in ascending order of $E$-value. Let $M$ be the total number
 of spectra used for evaluation and let $N_{E\le E_c}$ be the total number of false positives with
  $E$-values smaller than or equal to $E_c$. If the $E$-values reported are accurate, one expects to
 see that
\[
E_c = \frac{N_{E\le E_c}}{M}\; ,
\] 
subject to fluctuations due to finite sampling.

\begin{figure*}[htbp]
\begin{center}
\includegraphics[width=0.95\textwidth,angle=0]{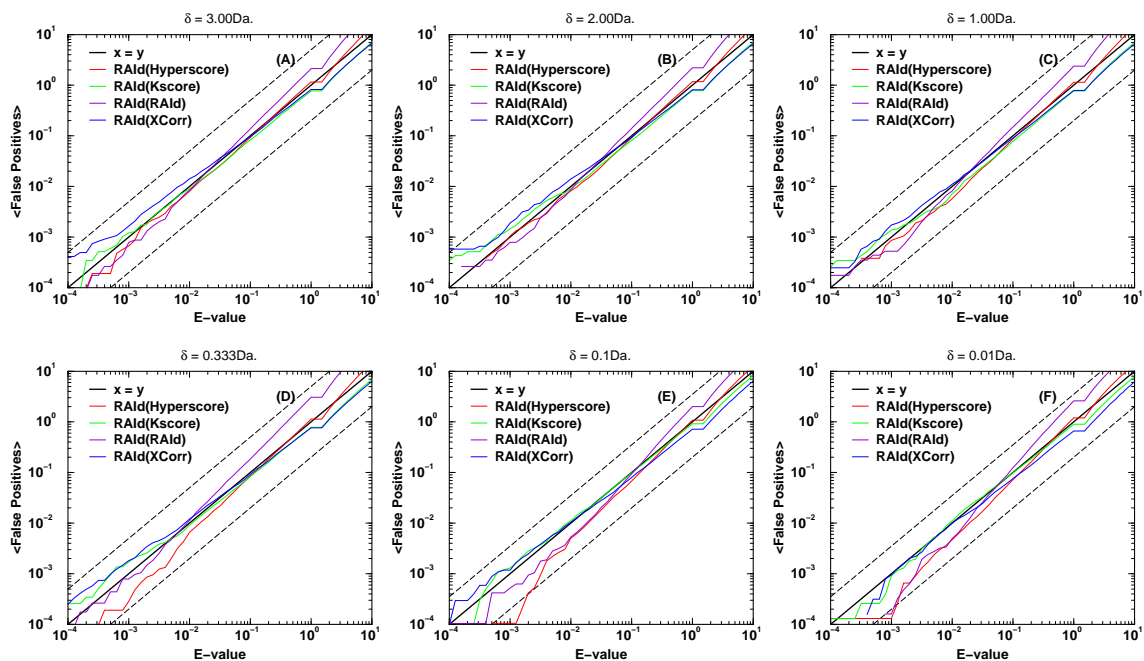}
\end{center}
\caption{E-value accuracy assessment. The agreement between the reported $E$-value and the textbook definition 
 is examined using centroid data (A1-A4 subsets of ISB data set). The random database size used is 500 MB. 
 The molecular weight range considered while searching the database is $[MW- \delta, MW + \delta]$. 
 In each panel, the dashed lines, corresponding to $x=5y$ and $x=y/5$, are used to provide a visual guide 
  regarding how close/off the experimental curves are from the theoretical curve.}
\label{E_accuracy_centroid}
\end{figure*}

Figures~\ref{E_accuracy_centroid} and S4 assess $E$-value accuracy when
 $E$-values are obtained from APP $P$-values. Figure~\ref{E_accuracy_centroid}
 displays, based on searching a random database of size 500MB, the measured average number of false positives
  as a function of the reported $E$-value. The six-panel figure demonstrates statistical stability against 
  allowed mass error. For parent ion mass of $2,000$ Da, what is displayed in Figure~\ref{E_accuracy_centroid} 
   covers the resolution range from $1,500$ ppm to $5$ ppm. 
   Figure~S5 displays the corresponding result for profile data. 
   The statistical stability shown is important since the 
   use of high resolution mass analyzers such as Orbitraps have gained popularity.   
    Figure S4, using the NCBI's nr database, examines the $E$-value accuracy when
     used in biological context. Since the biological database is not a collection of random peptides,
   the validity of statistical theory founded on random databases should be tested. As shown in Figure S5,   
   the same statistical robustness holds for both centroid and profile spectra while searching the
   biological protein database tested. 
 
 Both the centroid data set and profile
 data set are tryptic and are identical to the ones used in reference~\cite{GYWGetal_2007}. 
 The $E$-value for a peptide hit is obtained by multiplying that peptide hit's APP $P$-value
  by a numerical factor $N_d$, the number of qualified database peptides with similar masses. In terms of enumerating 
 qualified peptides, we employ the RAId\_DbS strategy. Specifically,  
 we further divide the qualified peptides into ones with correct and incorrect 
 N-terminal cleavages~\cite{RAId_DbS} and have separate counters for them.
 If a candidate peptide has correct
 N-terminal cleavage, its $N_d$ factor is the total number of database peptides with both
 correct N-terminal cleavages and with masses similar to that of the peptide considered;  
 otherwise, it will  have a considerably larger $N_d$ factor that counts 
 {\it all} database peptides with masses similar to that of the peptide considered.
  The protein database used is the NCBI's nr (same version as in reference~\cite{RAId_DbS}) 
 with identical cluster removal procedure~\cite{RAId_DbS}. As shown in Figure~\ref{E_accuracy_centroid} 
 and Figures~S4-S5, 
 the $E$-values reported by RAId\_aPS using the various scoring functions implemented are within a factor of five of the
 textbook definition. For any two scoring functions, if they 
 are independent, one may combine the statistics using eqs.~(\ref{E2P}-\ref{P2E})
  and the combined $E$-value should also follow the theoretical curves. 
  
 How well the combined $E$-values reported trace the theoretical line can be used as
 a measure of how independent these two scoring functions are, provided that each scoring function 
 already has $E$-value reported in agreement with the textbook definition. 
 As in reference~\cite{GWGFK_2008}, the combined $E$-value from any two methods
 in general shows a larger deviation from the textbook definition. This may be due to 
  correlations between search methods.    
  We are currently investigating the possibility of
   taking into account the search method correlation, which we suppose to be spectrum-specific too,
    while combining the statistics. 
We will incorporate the corrected statistics into RAId\_aPS 
if the investigation along this direction turns out to be fruitful.

\subsection*{Combine Database Search Results}
 The primary feature of RAId\_aPS is the ability to combine,  
 in a statistically sound way, search results from different scoring functions. 
 If the retrieval performance of each scoring function implemented is poor, then 
 even if one combines the search results, the final outcome might still be poor.  
 Below we assess the retrieval performance of each scoring function implemented 
   using the Receiver Operating Characteristic (ROC) curves. 

\subsubsection*{First assessment of scoring functions}
\begin{figure*}[htbp]
\begin{center}
\includegraphics[width=0.90\textwidth,angle=0]{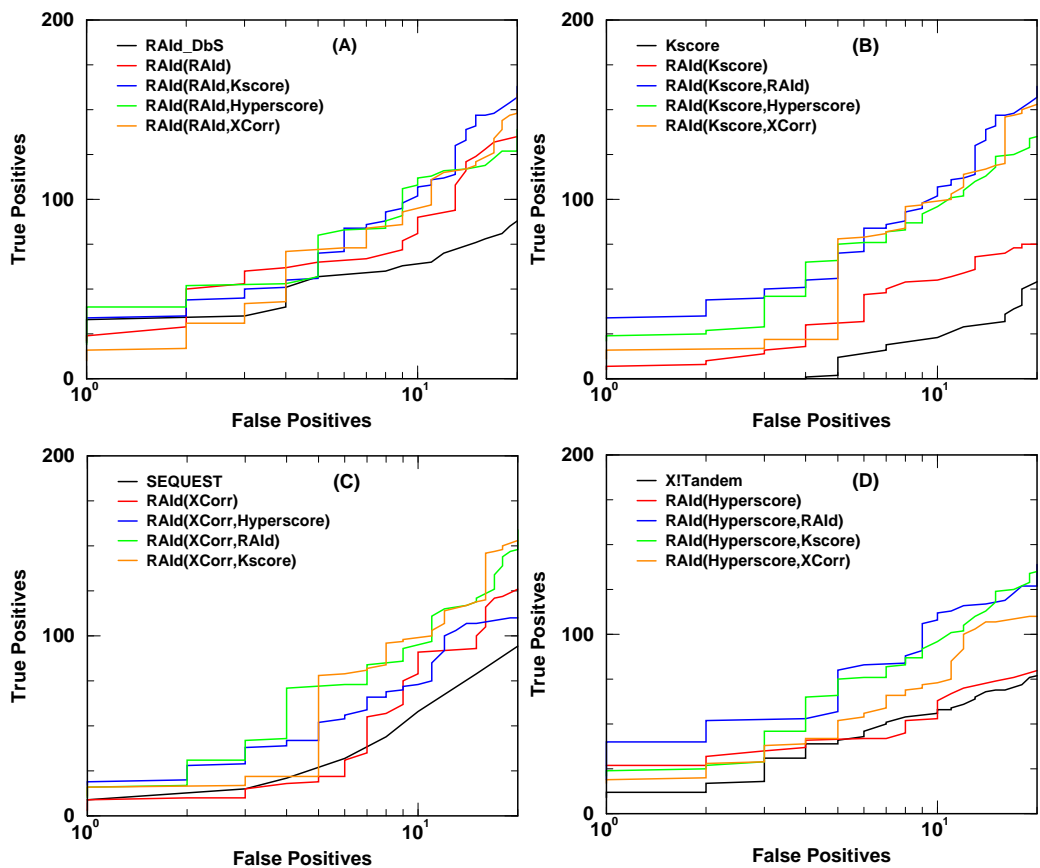}
\end{center}
\caption{ROC curves for the centroid data (A1-A4 of the ISB data set~\cite{KSNSetal_2002}). 
For each of the four scoring functions considered, 
 a set of ROC curves is shown. These ROC curves include the results from running the designated program 
 associated with that scoring function, the results from running RAId\_aPS in the database search mode,
 and the results from  combining with each of the three other scoring functions.  
  Panel (A) shows the results from RAId score, whose designated program
 is RAId\_DbS. Panel (B) displays the results from K-score, whose designated
 program is X!Tandem. Panel (C) exhibits the results from XCorr, which is
 mostly employed by SEQUEST. Panel (D) presents the results from Hyperscore,
 whose designated program is  also X!Tandem. Instead of using only XCorr (like RAId\_aPS), 
 SEQUEST first selects the top $500$ candidates using SP score. As shown in panel (C), for centroid data 
 there is an advantage to filtering candidates with the SP score. However, it is also seen that
 by combining XCorr with either RAId score or Hyperscore, equally good results can be attained without 
 introducing the SP score heuristics. 
}
\label{ROC_ISB}
\end{figure*}

\begin{figure*}[htbp]
\begin{center}
\includegraphics[width=0.90\textwidth,angle=0]{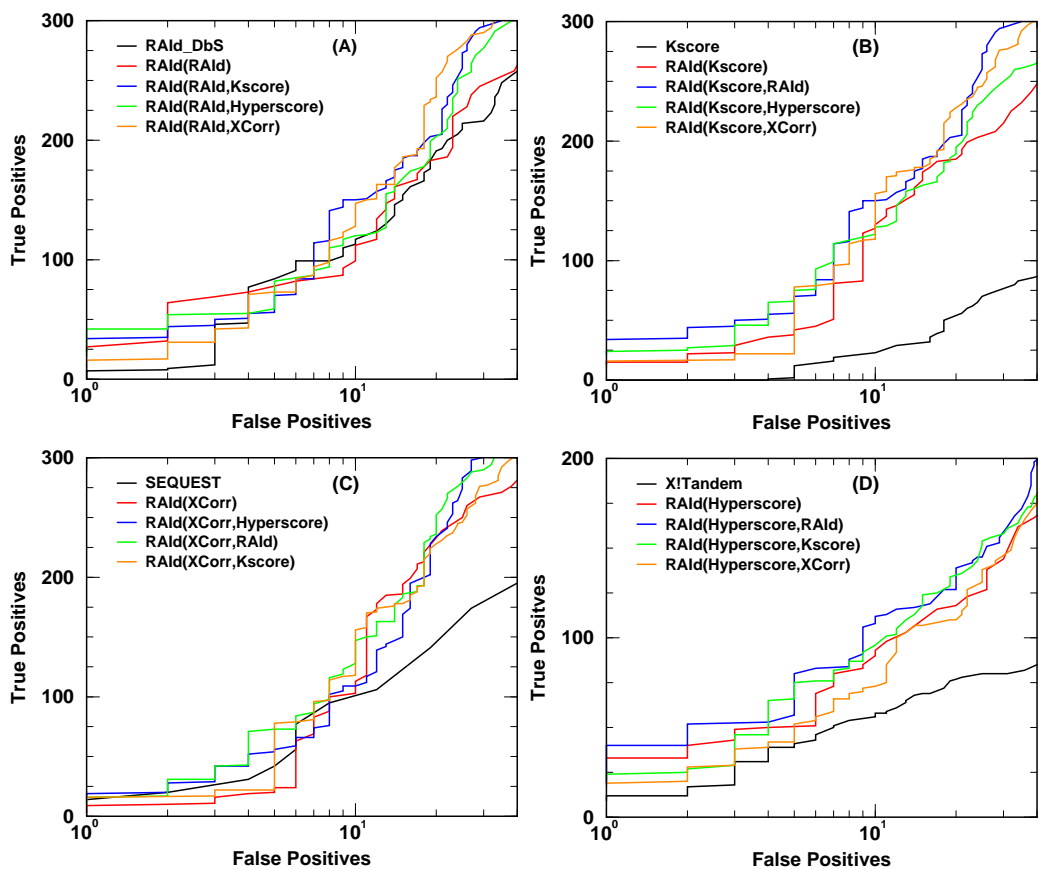}
\end{center}
\caption{ROC curves for the centroid data (A1-A4 of the ISB data set~\cite{KSNSetal_2002}) when considering only the best hit
 per spectrum. 
For each of the four scoring functions considered, 
 a set of ROC curves is shown. These ROC curves 
  include in the consideration only the best hit per spectrum from running the designated program 
 associated with that scoring function, the best hit per spectrum from running RAId\_aPS in the database search mode, 
 and the best hit per spectrum from combining with each of the three other scoring functions. 
 Panel (A) shows the results from RAId score, whose designated program
 is RAId\_DbS. Panel (B) displays the results from K-score, whose designated
 program is X!Tandem. Panel (C) exhibits the results from XCorr, which is
 mostly employed by SEQUEST. Panel (D) presents the results from Hyperscore,
 whose designated program is  also X!Tandem. Instead of using only XCorr (like RAId\_aPS), 
 SEQUEST first selects the top $500$ candidates using SP score. As shown in panel (C), for centroid data 
 there is advantage to filter candidates with the SP score. However, it is also seen that
 by combining XCorr with either RAId score or Hyperscore, equally good results can be attained without 
 introducing the SP score heuristics. 
}
\label{ROC_ISB_top}
\end{figure*}

Here we investigate the performance of the four implemented scoring functions
 --RAId score, K-score, XCorr, and Hyperscore-- each of which is 
 a standard scoring function, often employed with program-specific heuristics, for a known search program.  
 The retrieval efficiency is assessed using a centroid data  set (Figure~\ref{ROC_ISB}, ISB data set).
   Since many search methods report only
  one or very few candidate peptides per spectrum, we also include this type of 
  ROC curve (Figure~\ref{ROC_ISB_top}) where only the best hit per spectrum is taken from the search results.  
The performance of this {\it ad hoc} truncation apparently leads to better retrieval 
  at small number of false positives,
  indicating the existence of false hits whose evidence peaks are homologous to that 
 of the true positive(s) associated with a spectrum. We are currently investigating the impact of the
 existence of these types of false positives on the statistical significance assignment. The
 results will be reported in a separate publication.
 The corresponding plots when using a profile data set (NHLBI data set) are shown 
   respectively in Figure~S6 (similar to Figure~\ref{ROC_ISB}) and Figure~S7 (similar to Figure~\ref{ROC_ISB_top}).

\subsubsection*{Different ROC analysis}
When the true positive peptides are not known {\it a priori}, 
there exist various strategies in classifying hits into true or false positives when making a ROC plot.
 These strategies, unfortunately, will make a notable difference in retrieval assessment. 
 For example, in a cell lysate experiment of a certain organism, it is customary to estimate the 
 number of false positive hits by introducing a decoy database during the data analysis. The main idea there
 is to first sort the peptide hits according to their scores. Then for each decoy hit, one 
 assumes that there is just one corresponding false hit in the target database. 
 This strategy has been used extensively~\cite{SNA_2008}. ROC analyses done this way generally count   
 false positives, which are highly homologous to the target peptides,  
 towards true positives. This has two effects: an overcount of true positives and a undercount of false positives. As a
 consequence, the ROC curves will appear more impressive. To mimick this situation, we 
 used BLAST to find in the NCBI's nr database highly homologous proteins to the target proteins used in the experiment  
 and include those proteins in our true positive set. This strategy produces ROC curves shown as the 
  solid curves of Figure~S8. 
 When compared to Figure~\ref{ROC_ISB} and Figure~S6, the ROC curves produced by 
 this strategy seem much more impressive. 

 Not counting highly homologous proteins as false positives would probably be agreeable. However, counting those
 peptides/proteins as true positives could be exaggerating. Therefore one may use a slightly different strategy: removing 
 from consideration proteins homologous to the target proteins, which is called 
  the cluster removal strategy~\cite{RAId_DbS}. The dashed curves of 
 Figure~S9 are ROC curves obtained this way. 
 This strategy also produces slightly more impressive ROC curves than in Figure~\ref{ROC_ISB} and Figure~S6. 
  Apparently, this indicates the highly homologous false positive hits are the ones that 
 degrade the retrieval performance. Thus, it can be useful to remove those false positives from consideration. 
 Keeping only the best hit per spectrum turns out to be one way to achieve this goal.

\subsubsection*{Combining Multiple Scoring Functions}
 Since different scoring functions have different spectral filtering strategies, it is often
  advantageous to combine the search results from several scoring functions. RAId\_aPS provides
 a simple user interface, allowing users to select several 
scoring functions at a time. A example output when several scoring functions are selected is shown in
 Table~\ref{tab:comb_E_example}.

\begin{figure*}[htbp]
\begin{center}
\includegraphics[width=0.85\textwidth,angle=0]{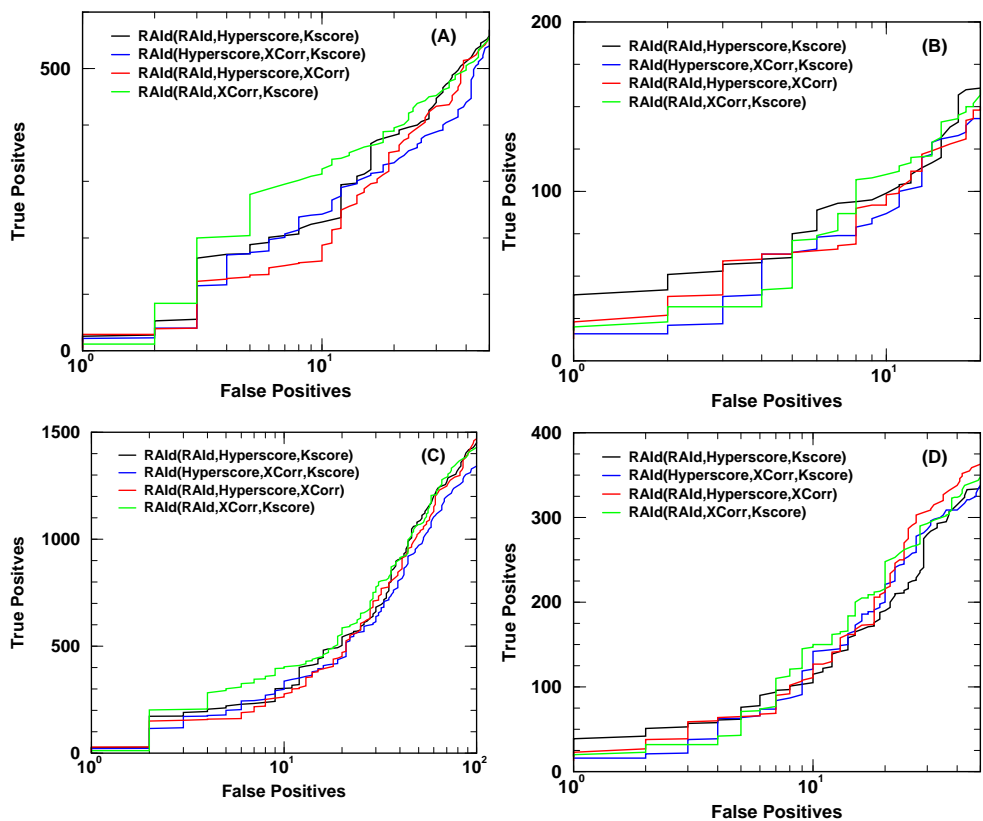}
\end{center}
\caption{
Illustration of RAId\_aPS performance when combining three different scoring functions. 
 Panel (A) shows the results from the profile data (NHLBI data set~\cite{GYWGetal_2007}), while panel (B) 
 exhibits the results from the centroid data (A1-A4 of the ISB data set~\cite{KSNSetal_2002}).   
 Panel (C) shows the results from the profile data but keeping only the best hit per spectrum, while
  panel (D) exhibits the results from the centroid data but keeping only the best hit per spectrum. 
}
\label{three_funcs}
\end{figure*}

Figure~\ref{three_funcs} illustrates the performance when RAId\_aPS combines three different scoring functions 
 in its database search mode. Panels (A) and (B) of Figure~\ref{three_funcs} should be compared 
  with Figure~\ref{ROC_ISB} and Figure~S6 respectively. The ROC curves obtained
 by combining three randomly chosen scoring functions indicate better performance  
 than individual scoring functions. Panels (C) and (D) should be compared with 
 Figures~\ref{ROC_ISB_top} and Figure~S7 respectively. The results in those plots 
 are obtained from
  keeping only the best hit per spectrum prior to further analysis. As shown in those plots,
   the ROC curves obtained by combining three randomly chosen scoring functions 
   indicate significantly better performance than individual scoring functions, except for the case of RAId\_DbS.

\subsection*{Other modes}
 Examples of using mode (iv) were already shown above.  
 We demonstrate here other features of RAId\_aPS to illustrate its versatility. 

\subsubsection*{Compute TNPP: mode (i)}
Given a parent ion mass, RAId\_aPS is also able to compute efficiently the
 TNPP associated with that molecular mass within a 
 user-specified mass error. The user interface for computing TNPP is self-explanatory. 
  One simply types in the molecular mass of interest, chooses a specific digesting enzyme or
  considers no enzymatic restriction by choosing ``no enzyme", and then presses the ``Submit a job"  button.
  If one wishes to change the default mass error tolerance, it can be done under the ``more parameter" toggle. 
  One may also elect to include PTMs or deselect certain amino acids from consideration,
    those choices are available under the ``Amino acids and PTMs" toggle. 
 When using search methods that do not have a theoretical model for the score distribution or when 
 the quality of the score model~\cite{RAId_DbS} is poor, one may
 wish to use a more conservative statistical significance assignment. In this case, a user may 
 set $1/{\rm TNPP}$ as the lower bound for the best $P$-value for any given parent ion mass. 
 This may help in preventing exaggerated/inappropriate statistical significance assignments.

\subsubsection*{Generate score histogram: mode (ii)}
\begin{figure}[htbp]
\begin{center}
\includegraphics[width=0.95\columnwidth,angle=0]{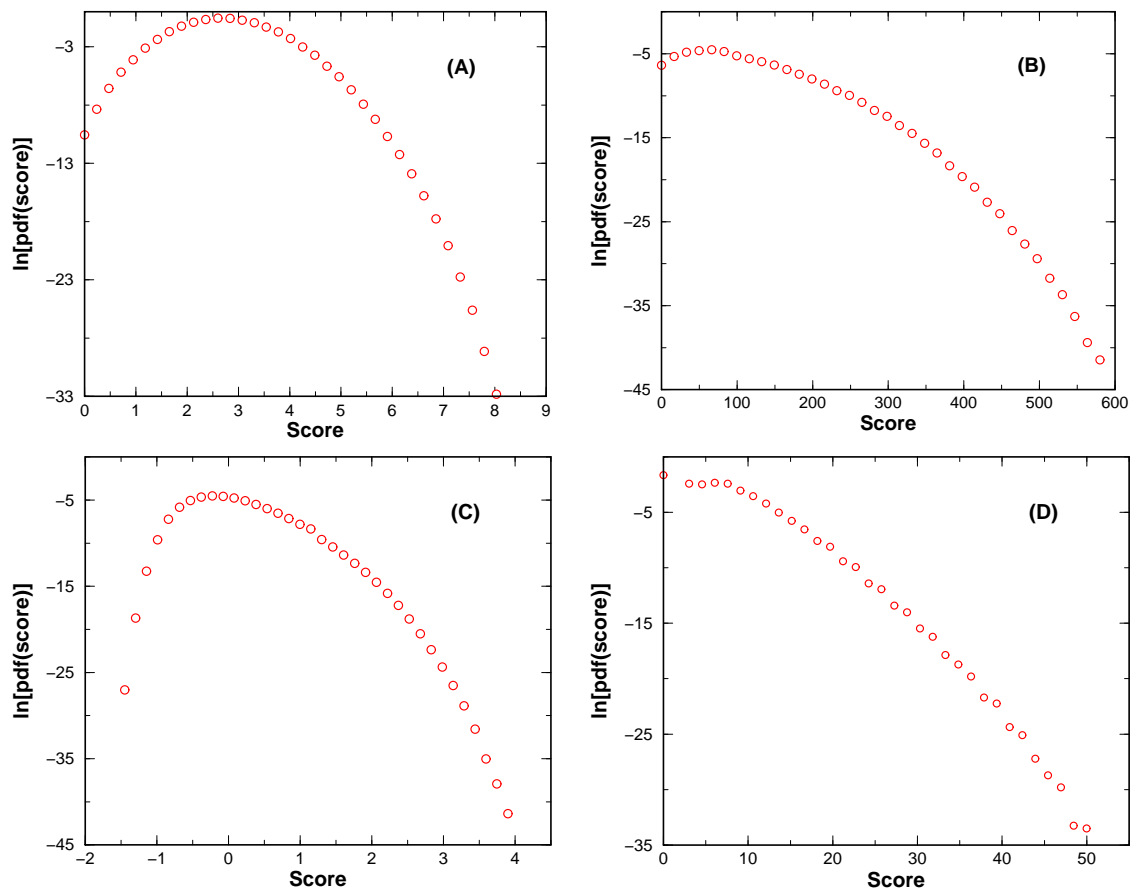}
\end{center}
\vspace*{-12pt}
\caption{\small Example score PDF (normalized histogram) output by RAId\_aPS. An MS$^2$ 
spectrum of parent ion mass $1640.80$ Da is queried  with default parameters, and the 
 resulting score PDF for RAId, K-score, XCorr, and Hyperscore are 
 shown respectively in panels A, B, C, and D.  
The number of APP within $\pm$ 3Da of parent ion mass is about $10^{19}$.}
\vspace*{-12pt}
\label{histogram_out}
\end{figure}

 Extraction of the statistical significance from a score distribution often 
 requires a model, be it theoretically derived or empirically assumed, 
 for the score distribution. 
One may test the robustness of a score model by examining how well the score model fits 
 the database search score histograms. When using search methods that have a score 
 model, one may first test how well the same score model
 applies when dealing with APP. If the score model loses stability, 
 this may indicate that the score model is not robust in general. 
 Given a query spectrum and a user-selected scoring function, RAId\_aPS can 
 be used to generate a score histogram of APP under the 
 selected scoring scheme. Using an example spectrum, Figure~\ref{histogram_out} 
 shows score histograms corresponding to the four scoring functions implemented in RAId\_aPS.

\subsubsection*{Reassign $\pmb{E}$-value : mode (iii)}
Statistical significance inference from RAId\_aPS only depends on the 
 total number of qualified peptides inside the database searched but is not dependent on 
 the peptide content inside the database. 
   This is because RAId\_aPS bases its statistics on  
 the (weighted) score histogram obtained from scoring APP. 
  As a consequence,  without going through the database search again, 
 RAId\_aPS can be used to reassign statistical significance to 
 a collection of candidate peptides. The candidate peptides may come from a flat list provided by the 
 user, or they can also come from the output files of various search engines. RAId\_aPS 
 allows users to upload the output files from SEQUEST, X!Tandem, and Mascot for statistical 
 significance reassignment.

\begin{figure*}[htbp]
\begin{center}
\includegraphics[width=0.89\textwidth,angle=0]{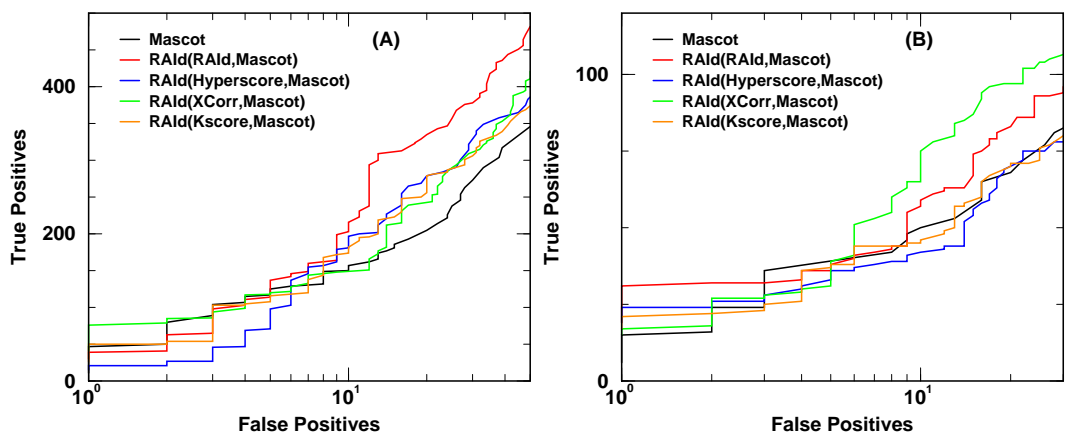}
\end{center}
\caption{Example of reanalyzing output files from other search engine by   
 combining with statistical significance assignment from RAId\_aPS. 
 In this example, we use the Mascot output files resulting from querying 
 profile spectra (panel (A), the NHLBI data set) and centroid spectra (panel (B), A1-A4 of the ISB data set~\cite{KSNSetal_2002})
 to the NCBI's nr database with proteins highly homologous to those that were present in the mixture removed.  
 Since each data set is from a known mixture of proteins, it is possible to remove the proteins homologous 
   to the true positives from the nr database. 
  We then combine the calibrated $E$-value~\cite{GYWGetal_2007} of Mascot with the $E$-value obtained from 
 RAId\_aPS when either RAId score, Hyperscore, K-score or XCorr is used. 
}
\label{Mascot_example}
\end{figure*}

Although scoring functions similar to XCorr, K-score and Hyperscore have
 been implemented in RAId\_aPS, other search engines' scoring functions might not be 
  suitable for score histogram construction using dynamic programming. In this case,
 the user may wish to compare the statistical significance reported by a search engine
 with what is reported by RAId\_aPS and even combine these reported significances.  
 As an example of this usage and to test RAId\_aPS's performance, 
 we use as queries $10,000$ profile spectra (the NHLBI data set) as well as
  $12,628$  centroid spectra (A1-A4 of the ISB data set), each produced from a known mixture of
 target proteins. Using Mascot as the search engine, 
 we searched in the NCBI's nr database with proteins highly homologous
 to the target proteins removed~\cite{RAId_DbS}. The output files were analyzed to produce ROC curves,
 the black solid curves in Figure~\ref{Mascot_example}. We then reanalyzed the candidate peptides' statistical
 significance by combining the statistical significance reported by Mascot with 
 that reported by RAId\_aPS using one additional scoring function. 
 For both profile and centroid spectra, when combined with either the RAId score, K-score, or XCorr , 
 one may obtain a retrieval performance that is comparable with or slightly better than that from Mascot alone
  (see Figure~\ref{Mascot_example}).

 Since all the implemented scoring functions are accessible from RAId\_aPS,
  one can score any new PTM peptide using any of the scoring functions available to
 RAId\_aPS even when the original program does not yet include the PTMs of
  interest. This way, annotated PTM found by RAId\_DbS~\cite{GYK_2008} 
 may be confirmed with other scoring functions in a natural manner
 and one may even combine the statistical significance as described below
 to increase the sensitivity in finding annotated PTMs and single amino acid
 polymorphisms (SAP).

\section*{Discussion}

In this section we will discuss another proposed use of the APP statistics
 in confidence assignment, \rev{remark on the effectiveness of combining search results
  using a different measure than ROC, propose} 
 avenues for improvement, and \rev{describe} future directions. 

When combined with database searches, the score histogram obtained by
 RAId\_aPS also provides two useful quantities. First, it gives us the best peptide score $S_{\rm APP}$ 
 among APP. Although we did not pursue this way, it has been advocated that the difference between $S_{\rm APP}$ 
  and the best database hit score per spectrum may serve as a statistical significance measure
 for the highest-scoring peptide hits found in the database~\cite{SNA_2008}. Second, the score histogram
 provides us with $N_s$, the (weighted) number of APP with score better than or 
 equal to $S$. This number $N_s$ may also be used in conjunction with the (relative) 
 difference between $S_{\rm APP}$ and the best database search score {\it per spectrum} 
 while constructing statistical significance measures other than $E$-value. 

\rev{A natural question to ask is: how much retrieval gain can one anticipate if one
 combines multiple scoring functions? Since FDR has been among the most popular metrics
  for assessing the performance, we briefly investigate this issue using FDR. 
  Employing a frequently used procedure~\cite{EG_2007A}, we used the reverse {\it Homo sapiens}
   protein database as the decoy database to estimate the number of false positives and hence the FDR, 
    by searching target database and decoy database separately for each query spectrum.  
 All $15$ possible combinations of the four scoring functions available in RAId\_aPS are 
  tested using the data set PRIDE\_Exp\_mzData\_Ac\_8421.xml (containing $15,916$ spectra), 
  downloaded from the PRoteomics IDEntifications (PRIDE) database (http:www.ebi.ac.ukprideppp2\_links.do).  
 The results are summarized in Table~\ref{tab:comb_method_FDR}   
 along with the average behavior associated with using one to four scoring functions. 
 Since it is known that performance of a search engine may vary when the data 
 to be analyzed changes\cite{JSHP_2009A}, we like to focus more on the average behavior rather than
  individual performance of a scoring function or any specific combination of scoring functions.
 Based on the average retrieval result of Table~\ref{tab:comb_method_FDR}, we first observe 
  that on average there is an overall retrieval increase at $0\% - 10\%$ FDR rates when one
   combine two scoring functions versus using only one scoring function.  
 We also note that there is an increase in retrieval performance at medium FDR rate when more 
   scoring functions are combined. However, at very low FDR rates, it seems that combining more than two
   scoring functions stop helping the retrieval. Apparently, 
   the performance boost does not continue indefinitely
  as more scoring functions are included. This is evidenced by an observable performance decline
   at low FDR rate when one combine all four scoring functions and compared to combine only three.
   The saturation of performance gain is reasonable if one takes into account the fact that
  most scoring functions seek similar evidences,  
 the scope covered by combining more scoring functions can't keep increasing indefinitely. 
}

By integrating existing annotated information into organismal databases, 
RAId\_DbS is now able to incorporate during its data analysis annotated information
 such as SAP, PTM, and their disease associations if they exist~\cite{GYK_2008}. 
 This feature enables users to identify/include known polymorphisms/modifications 
 in their searches without needing to blindly allow all possible SAPs and PTMs first and then
 post process to look up the literature/databases for explanations. 
 Since all the implemented scoring functions of RAId\_aPS are now within the same framework,
 we can let each plug-in scoring function incorporate in its scoring 
  the new SAP/PTM peptides. This way, annotated SAP/PTM found by RAId\_DbS 
 may be confirmed by other implemented scoring approaches in a natural manner
 and one may even combine the statistical significances as described earlier 
 to increase the sensitivity in finding annotated SAPs/PTMs.

In the near future, we also plan to include more scoring functions in RAId\_aPS if their presence would enhance the
 retrieval performance without sacrifice statistical accuracy. For example, we will investigate the effect of 
 a new scoring function, the compound Poisson. This is a natural way to incorporate intensity information
 into Poisson count statistics. The other scoring approach we will investigate is to deconvolute the 
 peptide length information. The reason to consider this alternative arises from the observation that
  many scoring functions introduce different heuristics to correct for the scores associated with 
 candidate peptides of different lengths. The purpose of these peptide length correction factors is 
 to balance the fact that longer peptides are likely to find more evidence peaks and thus the collected 
 evidence scores may require some length correction in order to make the comparison among peptides of
 various lengths impartial. If we group peptides of the same lengths and obtain statistical significance 
 separately for peptide candidates of each length, we no longer need to introduce any length correction
 factor. This approach is not feasible for regular database searches since the sample size of peptides of a fixed 
 length may be too small. For our APP scheme, however, we always have a large number of peptides 
  participating in our score histogram even if the peptide length is fixed. Therefore, the idea 
 of deconvoluting the peptide lengths becomes feasible for RAId\_aPS. 


\section*{Acknowledgments}
We thank Jimmy Eng for useful correspondence on the
spectral filtering strategy of SEQUEST's XCorr. 
We also thank the administrative group of the NIH Biowulf clusters, where all the 
computational tasks were carried out. 

\clearpage

\newpage
\section*{Figure Legends}

\clearpage

\clearpage

\clearpage

\clearpage

\clearpage 

\clearpage

\clearpage

\clearpage

\clearpage


\clearpage
\section*{Tables}

\begin{table*}[htbp]
\caption{ An output example of the combined {\it E-value} from RAId\_aPS. } \vspace*{0.1cm}
\begin{tabular}{|r|r|r|r|r|r|}\hline  
 & & & & & \\ [-0.6cm]
E\_comb & RAId \phantom{2} &  Hyperscore & XCorr\phantom{2}& K-score\phantom{l}& Peptide \phantom{peptidepeptide} \\ 
\hline
$4.93 {\rm e}{\small -24}$ & $1.69 {\rm e}{\small -13}$ &  $8.26 {\rm e}{\small -11}$   & $5.87 {\rm e}{\small -12}$  
& $7.99 {\rm e}{\small -13}$ &  NYQEAKDAFLGSFLYEYSR     \\ \hline
$1.43$ &     $379.00  $ &  $0.08$   & $453.00$      & $101.00$    &  APTSAGPWEKPTVEEALESGSR   \\ \hline
$1.85$ &     $28.50 $ &  $1.94$    & $9.01 $    & $0.15 $  &  LERMTQALALQAGSLEDGGPSR   \\ \hline
$3.38$ &     $13.60$ &  $0.30$   & $88.40$     & $4.32$   &  TEDQRPQLDPYQILGPTSSR   \\ \hline
$4.04$ &     $15.80$ &  $18.40$    & $0.38$     & $18.30$   &  NYKAKQGGLRFAHLLDQVSR   \\ \hline
$8.81$ &     $257.00  $ &  $1.48 $   & $1170.00$     & $1280.00$   &  DTPMLLYLNTHTALEQMRR    \\ \hline
$9.58$ &     $8.76 $ &  $1.66 $   & $353.00$      & $37.20$   &  EKTESSGQETTAKCDRASKSR   \\ \hline
$9.75$ &     $1.71 $ &  $8.15 $   & $82.80$     & $6.99$   &  LLAQQSLNQQYLNHPPPVSR   \\ \hline
$10.80$ &     $358.00  $ &  $1.95 $   & $311.00$      & $269.00$    &  IQHGQCAYTFILPEHDGNCR   \\ \hline
\end{tabular}
\label{tab:comb_E_example}
\end{table*}

\begin{table*}[htbp]

\caption{\rev{Example retrieval tests based on FDR. 
All $15$ possible combinations of the four scoring functions available in RAId\_aPS are shown 
 along with the average behavior associated with using one to four scoring functions. 
 The dataset  PRIDE\_Exp\_mzData\_Ac\_8421.xml is used. 
 The first column documents various combinations of scoring functions with the following 
  abbreviations: R for RAId, K for K-score, H for hyperscore, and  X for XCorr. 
 The rest of the columns display the number of peptides identified at the false positive rate
  specified at the top of the column.   
 The rows with bold characters indicate the average behavior of using a single (${\mathbf S}$) scoring function, 
 combining two (${\mathbf D}$) scoring functions, combining 
  three (${\mathbf T}$) scoring functions, and combining four (${\mathbf Q}$) scoring functions. 
  Within these rows, except the last one where only one combination possible, the standard deviation 
  associated with each average is shown inside the parentheses to the right of the average.}
} \vspace*{0.1cm}
\rev{
\begin{tabular}{|c|r|r|r|r|}\hline 
& & & & \\ [-0.6cm]
Combination  &   FDR cutoff  0\%   &  FDR  cutoff  2.5\%   &  FDR  cutoff   5.0\%  &  FDR  cutoff  10\% \\  \hline
R  & 377 \phantom{(3\hspace*{4pt}86)}& 822\phantom{(7\hspace*{17.5pt}8)}& 856\phantom{1\hspace*{23pt}1}& 948\phantom{1\hspace*{22.5pt}1}\\  \hline
K  & 83  \phantom{(3\hspace*{4pt}86)}& 709\phantom{(7\hspace*{17.5pt}8)}& 790\phantom{1\hspace*{23pt}1}& 977\phantom{1\hspace*{22.5pt}1} \\  \hline
H  & 568 \phantom{(3\hspace*{4pt}86)}& 775\phantom{(7\hspace*{17.5pt}8)}& 849\phantom{1\hspace*{23pt}1}& 908\phantom{1\hspace*{22.5pt}1}\\   \hline
X  & 467 \phantom{(3\hspace*{4pt}86)}& 821\phantom{(7\hspace*{17.5pt}8)}& 885\phantom{1\hspace*{23pt}1}& 996\phantom{1\hspace*{22.5pt}1}\\   \hline
$\overline{\mathbf S} (\sigma_S)$ & {\bf 373 (182)}  & {\bf 781 (57)}\phantom{(1)}\hspace*{-5pt}  & {\bf 845 (34)}\phantom{15}\hspace*{-5pt}  & {\bf 957 (39)}\phantom{15}\hspace*{-5pt} \\   \hline 
RK & 485 \phantom{(3\hspace*{4pt}86)}&956\phantom{(7\hspace*{17.5pt}8)}&  1127\phantom{1\hspace*{23pt}1}& 1654\phantom{1\hspace*{22.5pt}1}\\   \hline
RH & 925 \phantom{(3\hspace*{4pt}86)}&1143\phantom{(7\hspace*{17.5pt}8)}& 1599\phantom{1\hspace*{23pt}1}&2375\phantom{1\hspace*{22.5pt}1}\\   \hline
RX & 871 \phantom{(3\hspace*{4pt}86)}&1024\phantom{(7\hspace*{17.5pt}8)}& 1140\phantom{1\hspace*{23pt}1}&1574\phantom{1\hspace*{22.5pt}1}\\   \hline
KH & 528 \phantom{(3\hspace*{4pt}86)}&1019\phantom{(7\hspace*{17.5pt}8)}& 1210\phantom{1\hspace*{23pt}1}&1679\phantom{1\hspace*{22.5pt}1}\\   \hline
KX & 588 \phantom{(3\hspace*{4pt}86)}&860\phantom{(7\hspace*{17.5pt}8)}&964\phantom{1\hspace*{23pt}1}&  1146\phantom{1\hspace*{22.5pt}1}\\  \hline
HX & 895 \phantom{(3\hspace*{4pt}86)}& 1064\phantom{(7\hspace*{17.5pt}8)}&1205\phantom{1\hspace*{23pt}1}&1532\phantom{1\hspace*{22.5pt}1}\\  \hline
$\overline{\mathbf D} (\sigma_D)$ & {\bf 715 (186)}  & {\bf 1011 (87)}\phantom{(1)}\hspace*{-5pt} & {\bf 1207 (196)}\phantom{1}\hspace*{-5pt}  & {\bf 1660 (365)}\phantom{1}\hspace*{-5pt}    \\  \hline 
RKH & 485 \phantom{(3\hspace*{4pt}86)}&849\phantom{(7\hspace*{17.5pt}8)}&2689\phantom{1\hspace*{23pt}1}& 5328\phantom{1\hspace*{22.5pt}1}\\   \hline
RKX & 474 \phantom{(3\hspace*{4pt}86)}&792\phantom{(7\hspace*{17.5pt}8)}& 1074\phantom{1\hspace*{23pt}1}& 2425\phantom{1\hspace*{22.5pt}1}\\  \hline
RHX &725 \phantom{(3\hspace*{4pt}86)}& 867\phantom{(7\hspace*{17.5pt}8)}& 1942\phantom{1\hspace*{23pt}1}& 4795\phantom{1\hspace*{22.5pt}1}\\  \hline
KHX & 443 \phantom{(3\hspace*{4pt}86)}& 658\phantom{(7\hspace*{17.5pt}8)}& 910\phantom{1\hspace*{23pt}1}& 1691\phantom{1\hspace*{22pt}1}\\ \hline
$\overline{\mathbf T} (\sigma_T)$& {\bf 531 (116)} &  {\bf 791 (86)}\phantom{(1)}\hspace*{-5pt}  & {\bf 1653 (716)}\phantom{1}\hspace*{-5pt} & {\bf 3559 (1537)}\hspace*{-5pt}         \\  \hline
RKHX (${\mathbf Q}$) & {\bf 332} \phantom{(3\hspace*{4pt}86)}& {\bf 662}\phantom{(7\hspace*{17.5pt}8)}& {\bf 1336}\phantom{1\hspace*{23pt}1}& {\bf 4148}\phantom{1\hspace*{22.5pt}1}\\ \hline
\end{tabular}
\label{tab:comb_method_FDR}
}
\end{table*}

\end{document}